\documentclass[useAMS,usenatbib]{mn2e}
\usepackage{graphicx}
\usepackage{natbib}
\usepackage{hyperref}
\usepackage{aas_macros}
\usepackage{amsmath}

\title[Red galaxies with pseudo-bulges in SDSS]{Red galaxies with pseudo-bulges in the SDSS: closer 
to disk galaxies or to classical bulges?}
\author[B. Ribeiro et al.]{B. Ribeiro$^{1}$\thanks{E-mail:
bruno.ribeiro@lam.fr} and C. Lobo$^{2,3}$ and S. Ant\'on$^{4,5}$ and J.M. Gomes$^{3}$ and P. Papaderos$^{3}$\\
$^{1}$ Aix Marseille Universit\'e, CNRS, LAM (Laboratoire d'Astrophysique de Marseille) UMR 7326, 13388, Marseille, France\\
$^{2}$ Departamento de F\'isica e Astronomia, Faculdade de Ci\^encias,
Universidade do Porto, Rua do Campo Alegre 687, PT4169-007 Porto, Portugal\\
$^{3}$ Instituto de Astrof\'isica e Ci\^encias do Espa\c{c}o, Universidade do Porto, CAUP, Rua das Estrelas, PT4150-762 Porto, Portugal\\
$^{4}$ Instituto de Astrof\'isica de Anadaluc\'ia (CSIC), Glorieta de la Astronomia, s/n  18008, Granada, Spain\\
$^{5}$ Instituto de Astrof\'isica e Ci\^encias do Espa\c{c}o, Universidade de Lisboa, Faculdade de Ci\^encias, Campo Grande, PT1749-016 Lisboa, Portugal}

\begin{document}

\date{ }

\pagerange{\pageref{firstpage}--\pageref{lastpage}} \pubyear{2015}

\maketitle

\label{firstpage}

\begin{abstract}
Pseudo-bulges are expected to markedly differ from classical, 
quasi-monolithically forming bulges in their star formation history (SFH) 
and chemical abundance patterns.
To test this simple expectation, we carry out a comparative structural 
and spectral synthesis analysis of 106 red, massive galaxies issued from the SDSS, subdivided into 
bulgeless, pseudo-bulge and classical bulge galaxies according 
to their photometric characteristics, and further obeying a specific selection to minimize uncertainties in the analysis and ensure an unbiased derivation 
and comparison of SFHs.
Our 2D photometry analysis suggests that disks underlying pseudo-bulges typically 
have larger exponential scale lengths than bulgeless galaxies, despite similar integral 
disk luminosities. 
Spectral synthesis models of the stellar emission within the 3$\arcsec$ 
SDSS fiber aperture reveal a clear segregation of bulgeless and pseudo-bulge 
galaxies from classical bulges on the luminosity-weighted planes of age-metallicity 
 and mass-metallicity, though a large dispersion is observed within the two former classes. 
The secular growth of pseudo-bulges is also reflected upon their cumulative stellar mass
as a function of time, which is shallower than that for classical bulges.
Such results suggest that the centers of bulgeless and pseudo-bulge galaxies 
substantially differ from those of bulgy galaxies with respect 
to their SFH and chemical enrichment history, which likely points to different formation/assembly mechanisms.
\end{abstract}


\section{Introduction}\label{sec_intro}

The study of galaxy bulges has gained significant momentum lately, much due to the 
realization that the nature and formation history of these structural components of galaxies 
is far more intricate and diverse than what was considered to be some years ago.
 It has been suggested that some bulges - named pseudo-bulges  - 
emerged through a prolonged star-formation process that is fed through disk material 
in the course of the secular galaxy evolution, in contrast to classical bulges, which are 
thought to assemble through quasi-monolithic gas collapse and/or mergers early on 
\citep[for a reveiw see][see also references therein]{kk04} .
Due to their distinct formation history, pseudo-bulges are thus expected to appreciably
differ from classical bulges in their stellar content (age, metallicity, dynamics, morphology 
and structure) and current star formation rate (SFR).
Reproducing the formation of these structures within the hierarchical merging 
paradigm of galaxy formation and evolution is challenging 
\citep[see, e.g.,][and references therein]{okamoto13} yet of considerable relevance, as 
nearby disk galaxies frequently host structures that can be considered 
as pseudo-bulges  \citep[e.g.][]{kor10,simard11,fisher11,bennert14}. 

\citet{kk04} list several characteristics that can be used to identify
pseudo-bulges, one of them being a surface brightness profile (SBP) 
that is shallower than the one displayed by classical bulges. 
A fit with a S\'ersic law to pseudo-bulges is thus expected to yield 
a lower S\'ersic index $\eta$ (generally $<$2, i.e. closer to the value $\approx$1 
that is typical for disks) instead of the high $\eta$ (3--6) that is characteristic 
of massive galaxy spheroids, and closer to the de Vaucouleurs law.
A significant degree of rotational support and diskiness are regarded as 
further signatures of pseudo-bulges.
Their distinctiveness in terms of structural, morphological and kinematic 
properties, relatively to what one historically calls a bulge, is now relatively 
well established and widely used to identify them
\citep[e.g.][]{fishdrory08,gadotti09}.

Finer photometric details, harder to observe and hardly hinted on SBPs, may further help identifying pseudo-bulges. Unsharp masking techniques applied to images often reveal underlying structural features (spiral arms, bars, rings, small scale disks) considered to be typical of pseudo-bulges (e.g., \citealt{kk04} and references therein). These sometimes co-exist with a classical bulge (see e.g. \citealt{erwin15}) and are thought to be built up from the disk material, also based on kinematics (see also e.g. \citealt{keh12} for detection of such structures in early-type galaxies possessing a warm inter-stellar medium).
This lends further support to a ``gentler'' formation mechanism that is associated 
with internal gravitational disk instabilities and prolonged star-forming activity 
that could be sustained by gas accretion from the galaxy environment, rather than quick 
growth through merging and the ensuing violent relaxation of the stellar component. 
Mergers were - at least until recently - commonly expected to destroy disks 
in general, and likely also the faint spiral signatures described above, rather giving rise to a 
classical bulge (eg \citealt{naab06,hopkins10}). 
Nevertheless, there is also some theoretical and observational evidence for the survival 
of disks after mergers, or the existence of pseudo-bulge structures in merger remnants 
\citep[][and references therein]{kesel12,ueda14,quer14}. Alternatively, \citet{guedes13} 
propose a scenario where mergers possibly induce rapid growth of pseudo-bulges 
out of disk material in the early stages of galaxy assembly.
In this scenario the structure that we recognize as a pseudo-bulge today results from the complex evolution 
of a stellar bar that suffers repeated stages of formation and dissolution along the 
galaxy's evolution. Indeed, bars are thought to drive physical processes that lead to  
pseudo-bulge formation (\citealt{combes90,kk04,kubryk13}; see also \citealt{mendezabreu14} 
and references therein) traditionally in a secular way. But other authors minimize the contribution provided by secular bar instabilities to the assembly of pseudo-bulges in favour of physical mechanisms occurring at much smaller timescales like e.g. high-redshift starbursts \citep{okamoto13} or minor mergers \citep[][]{eliche11}. 

Given the plethora of possibilities, the existence of specific structural components 
such as embedded bars should not, in principle, be a requirement for identifying a pseudo-bulge.
And, somewhat in tandem, the timescale for the formation of the
pseudo-bulge (thus the dominant mechanism leading to its formation, provided that there 
is a unique one) is not yet pinned down either; neither by simulations nor even by observations. 
Still on the latter, a recent analysis based on the colors of a sample of isolated galaxies
seems to discard as well the need for long timescales for the formation of
pseudo-bulges: \citet{fer14} finds that the majority of their pseudo-bulges
show colors compatible with the red sequence of early-type galaxies, thus
advocating ``an early formation epoch and not much subsequent growth''.
However, the family of pseudo-bulges might be far more diverse, with the old entities 
described \citep{fer14} constituting merely a sub-branch of it. 
In fact, several previous studies document the presence of younger, bluer stellar populations 
and significant SFR in pseudo-bulges (e.g., \citealt{morelli08,fdf09,gadotti09,zhao12} 
and references therein), consistent with a time-continued build-up process, 
in contrast to that undergone by classical bulges.

All these works show that the defining characteristics - in particular in what concerns
stellar populations and morphological substructure - and origin of pseudo-bulges
are far from being well understood. 
A thorough census and analysis of pseudo-bulges is thus imperative to allow for advances 
to be made in this field. 
And, starting by the nearby systems, which can be more easily observed, is an obvious choice.

The aim of this study is to contribute to our understanding of such
structural entities, using a sample of galaxies selected on stringent criteria, and
employing a stellar population synthesis (SPS) approach,
which has not yet been
very much explored and can provide relevant constraints. 
Indeed, stellar ages, metallicities, star formation histories (SFHs) and other parameters 
inferred from SPS can give insight into the formation mechanism of pseudo-bulges and help addressing 
whether it is distinct from that of classical ones. 
Our sample comprises galaxies with little or absent ongoing star formation,
mainly to avoid the difficulties in the analyses of their SBPs 
that are due to light contamination by H{\sc ii} regions and 
associated features (e.g., nebular emission, dust lanes), that complicate 
the determination of the S\'ersic index $\eta$ in SBP fitting. 
This selection is described in Sect. \ref{sec_sample}, where we also detail 
the criteria adopted to divide the sample into different structural
classes and identify pseudo-bulge hosts. 
With this deliberate bias in star-forming activity (and color), the SPS should also be more reliable (due to faint emission lines and insignificant continuum nebular emission). We thus expect that, 
if any difference in the SFHs and stellar metallicities between pseudo-bulges and classical ones 
is still uncovered, then this should provide a robust indication. 
In Section 3 we give a brief explanation
of the method used to infer the main physical properties of the analyzed pseudo-bulges 
through spectral modeling of their SDSS spectra.
In Section 4 we provide and discuss the main results from our analysis. Finally, in Section 5 we summarize the main
points to be taken from this work.

Throughout the paper we use WMAP7 cosmology: H$_0$=71 km s$^{-1}$
Mpc$^{-1}$, $\Omega_m$=0.27 and $\Omega_\Lambda$=0.73 \citep{wmap7}.

\section{The sample - selection criteria and definition of structural classes}\label{sec_sample}

This study is largely motivated by and builds upon our previous work in 
\citet{coelho13} that focused on a sample of massive red galaxies selected
from the same NYU-VAGC
catalogue of \citet{blanton05} to ascertain the frequency of
active galactic nuclei (AGN) hosted by quiescent galaxies with a
negligible or absent bulge.
Since the \citet{coelho13} sample was restricted to low S\'ersic indices ($\eta <1.5$), we have extended it to include as well galaxies where the bulge has a significant contribution to the total light of the galaxy.
We thus select SDSS DR7 galaxies from the same NYU-VAGC
catalogue of \citet{blanton05} that gathers photometric and structural
parameters for all SDSS galaxies having spectroscopic data. All objects obey the following requirements:

\begin{itemize}
\item[-] redshift in the range $0.02 < z < 0.06$;
\item[-] galaxy stellar mass $M_* > 10^{10} M_\odot$, computed as in
  \citet{bell08} from the r-band SDSS luminosity and g-r color, and assuming
a \citet{chab03} initial mass function;
\item[-] color index $(g-r)$ typical of red galaxies:
 $(g-r) > 0.57 + 0.0575 ~log (M_*/10^8M\odot)$  \citep{bell08};
\item[-] inclination cut equivalent to $i \la 60^o$ to exclude edge-on systems,
  prone to biases in the photometric and SPS analysis due to dust extinction \citep[see][]{coelho13}.
\end{itemize}

These criteria yield a sample of galaxies with expected little star formation activity.
On top of the above, we also carried out a visual inspection of the 
SDSS multi-waveband images to discard galaxies with features obviously  
related to star formation, 
close-by or overlapping objects, as well as galaxies with interaction-induced distortions and 
complex morphologies 
(i.e., star-forming rings, tails and bridges, prominent dust lanes)
, since all these could compromise 
a robust estimate of the galaxies' structural properties, in particular in what regards the bulge. 
This second selection stage was also meant to minimize the number of obscured objects, 
which indeed was found to be the case in the subsequent analysis (Sect. \ref{subsec:EmLines}).

We expect this strategy to greatly simplify the morphological analysis and render it more
robust, while introducing no bias in what concerns the main
objective of this work: to investigate a possible significant difference 
in the stellar ages, metallicities and other properties of pseudo-bulges 
when compared with classical bulges.
More specifically, our main goal is to explore trends with regard to the 
above quantities or plainly to verify whether there are obvious contradictions 
to a scenario of different formation mechanisms/timescales between bulges 
and pseudo-bulges.

The adopted redshift interval and mass cutoff aim at obtaining a compromise of having enough objects in the sample, a good spatial coverage in photometry (for inspecting galaxy images and deriving significant SBPs), adequate spectroscopic sampling of their central regions and no particular correlations of the derived parameters with galaxy mass (due to the exclusion of low-mass objects).

The above selection criteria imposed a limit on massive, almost face-on,
pure disks with red colors of about 35 (for the redshift range in
question). The advantage of having a relatively small sample\footnotemark{ }
 \footnotetext{Because red late-type galaxies
  are rare \citep[e.g.,][]{bam09,masters10,coelho13}, a consequence of
  adopting our color selection is to drastically reduce the number of
  objects of later types, so our aim cannot possibly be to compile
  large samples that offer statistical robustness to the results.} 
is that its size allows doing a dedicated analysis of each object.
The approximately same number of galaxies was sought
(in the same volume) to populate the other classes, resulting in a
total of 106 objects. Their morphological classification is discussed
in the next sub-section.

\subsection{Structural analysis of the sample}\label{sec_galfit}

One of the commonly used ways to distinguish between bulge dominated
galaxies and disk dominated ones, and assess the significance of the
bulge when it is present, is to fit a \cite{sersic68} profile
\begin{equation}
	\centering
	\Sigma (r) = \Sigma_e \exp[-\kappa(r/r_{e})^{1/\eta}-1] \label{sersic}
\end{equation}
where the index $\eta$ describes the shape of the SBP, $r_e$
is the effective radius, $\Sigma_e$ is the surface brightness
at radius $r_e$ and $\kappa$ is a parameter coupled to $\eta$ \citep{ciotti99},
such as to ensure that half of the total flux is enclosed within $r_e$. 
A S\'ersic index of $\eta=1$ corresponds to a pure exponential profile that is typical 
for galactic disks, whereas $\eta=4$ corresponds to the de Vaucouleurs profile 
associated to elliptical galaxies and low-mass galaxy spheroids.

Though values of $\eta$ were already available for our galaxies in the
NYU-VAGC catalog, these were obtained through the fitting of equation
\ref{sersic} to the azimuthally averaged radial profile of each
galaxy, convolved with the estimated seeing \citep{blanton05}. 
By definition, such a profile determination and fitting technique 
collapses a 2D image into a 1D SBP.
Although this transformation is straightforward in its practical application,
and permits a standardized quantification of galaxy structural properties, it obviously entails loss of information.
For example, one can infer from decomposition of an SBP the total magnitude and 
mean isophotal radius of a structural component (e.g., a bar) but important  
morphological properties (e.g., position angle, ellipticity and boxiness/diskiness)
are irretrievable from a 1D SBP.
2D surface photometry techniques, now commonly used, can, in principle, preserve 
much of the 2D information in a galaxy image, and it has been claimed on the basis of  
simulations that they can better recover galaxy structural components, when these 
follow simple empirical laws \citep{wadadekar99,guo09}.

We performed 2D photometry modeling with GALFIT\footnotemark{ }
\footnotetext{http://users.obs.carnegiescience.edu/peng/work/galfit/galfit.html} \citep[][version 3.0]{peng02,peng10} 
to the r-band SDSS images, with the goal of analysing in more detail the structural parameters of the selected galaxies and assigning them to different morphological subsamples. To each of the 106 galaxies we carried out a fit with a single S\'ersic (equation \ref{sersic}) to the entire galaxy. Two types of galaxies (60\% of the whole sample) were well fit using this strategy: those described by a S\'ersic law with $\eta = 1$ (38 galaxies, hereafter named disks or bulgeless galaxies) and another group well modeled by a S\'ersic function with $\eta > 3$ (26 galaxies, for which no disk component was required - what we will call the bulgy galaxies).
 However, results were inadequate for the remaining 40\% of the sample (42 galaxies) since we identified significant signal in the residuals image. The SBP of these objects was thus refit with a multi-component model consisting of an exponential disk plus a S\'ersic profile - translated by the following equation:
\begin{equation}
\centering
	\Sigma (r) = \Sigma_{e,d}\exp[-\kappa(r/r_{e,d})-1] + \Sigma_{e,b}\exp[-\kappa(r/r_{e,b})^{1/\eta_b}-1].  
\label{multisersic}
\end{equation}

\noindent where subscripts $d$ and $b$ refer to disk and bulge components, respectively. 
This allowed to obtain good results for the whole sample (residuals after model subtraction amount to less than 10\% of the galaxy flux). No additional constraint on the profiles, aside from the fixed S\'ersic
index for the disk component (when required) in equation \ref{multisersic}, was imposed on all
fits.
The sky component was fitted locally and simultaneously using an
image with a constant value close to that reported from the SDSS. The
errors associated with each pixel are computed internally by GALFIT
from a combination of the rms of the sky regions and the pixel signal
assuming a poissonian distribution of the noise
\footnotemark{}\footnotetext{For a detailed description on how this is
  done please refer to
  \href{http://users.obs.carnegiescience.edu/peng/work/galfit/CHI2.html}{http://users.obs.carnegiescience.edu/peng/work/galfit/CHI2.html}}.

 Consistency tests
led us to confirm the goodness of this strategy where we fit a fixed $\eta=1$
S\'ersic law plus a free-$\eta$ S\'ersic profile  (i.e., equation \ref{multisersic}) to all galaxies\footnotemark{ }
\footnotetext{Note that , in practice, equation \ref{multisersic} was fit to all galaxies, considering that the second term is null in the case of bulgeless galaxies whereas it's the first term that is canceled for pure spheroids.}
since it turned out to
be more robust against possible variations in the other structural
parameters, and to yield more physically meaningful results.
Additionally, fitting a pure exponential disk to the
extended regions of the galaxy allows for a consistent measurement of
the light in excess to it, which can plausibly be attributed to the
bulge emission. This more robust approach is also justified by an
important goal of this study, namely to quantify the significance and
luminosity contribution of the bulge even for almost bulgeless objects
showing a merely minor luminosity surface brightness enhancement at
the centers of their disk 
though having no conspicuous sub-structure
(like bars, rings, etc). 
\begin{figure}
\centering
\includegraphics[width=\linewidth]{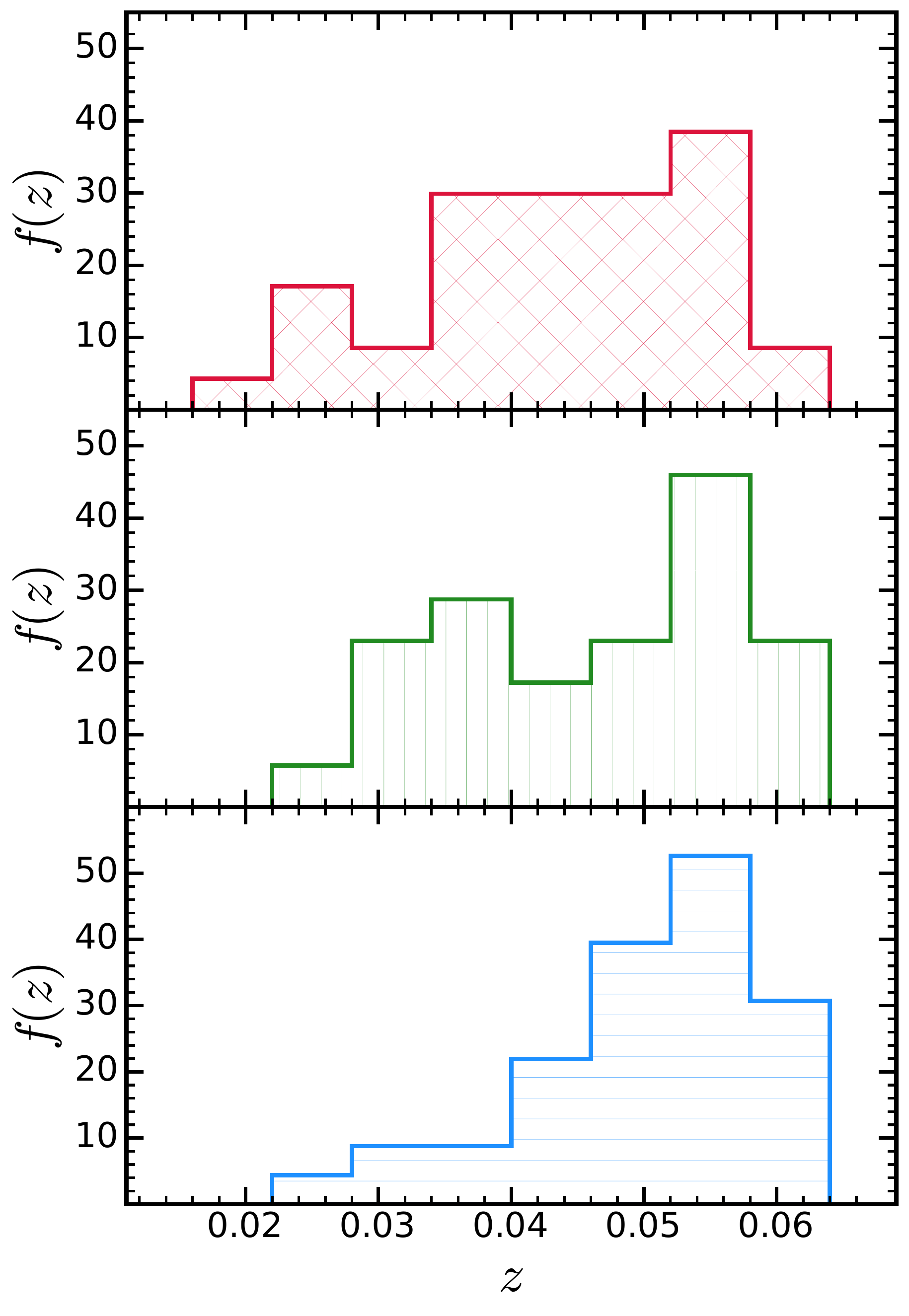}
\caption{\label{fig_historedshift} Redshift distribution of the different sub-samples. Top: intermediate+bulgy (these two classes are merged for the purpose of SPS studies, as detailed in section \ref{sec:synth}); Middle: pseudo-bulge; Bottom: bulgeless. f(z) represents the number of galaxies divided by the integral of the distribution, for normalization.}
\end{figure}

\begin{figure*}
\centering
\includegraphics[height=0.9\textheight]{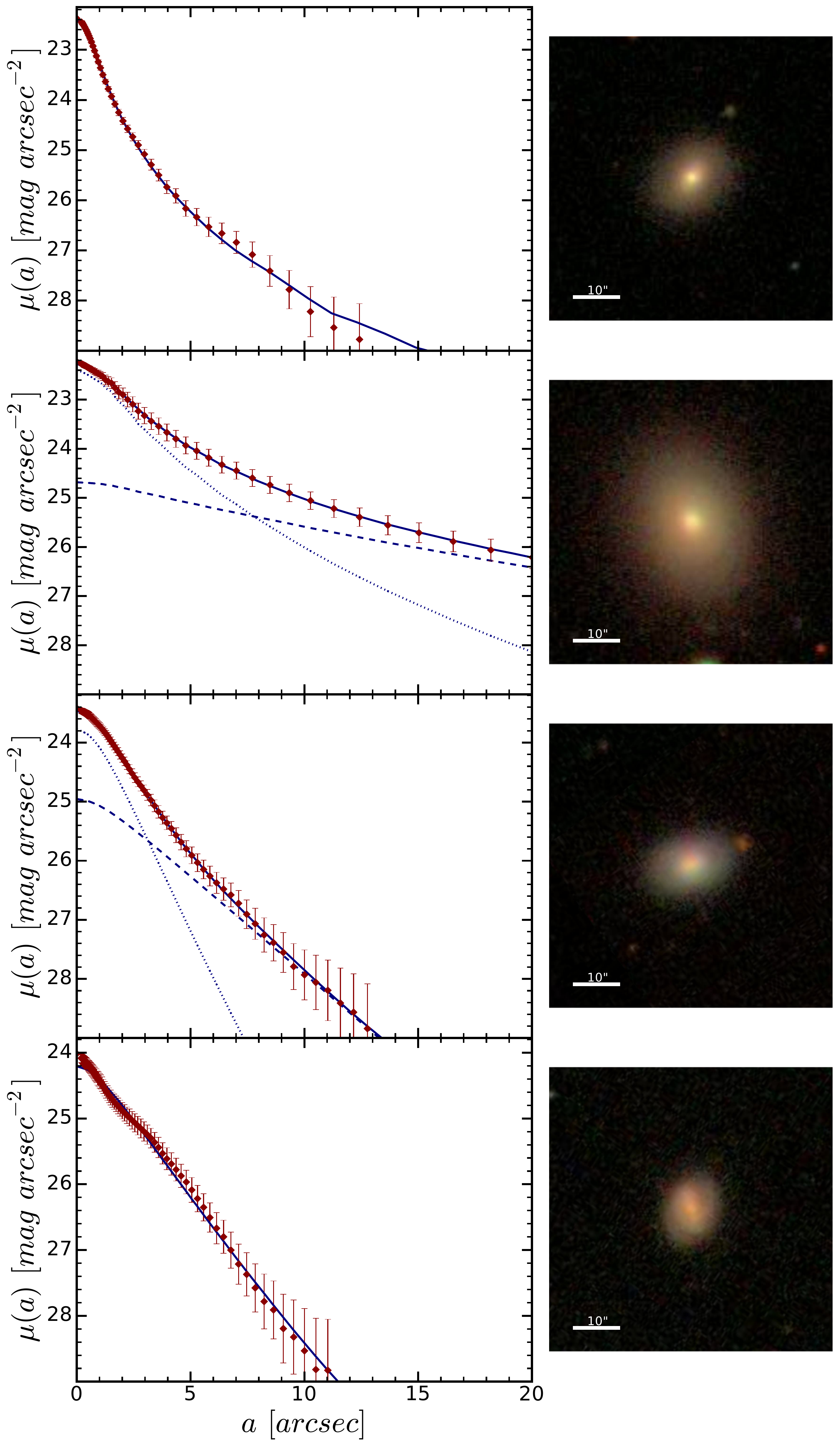}
\caption{\label{fig_galaxyprofiles} Left: typical examples of 1D surface brightness profiles (SBP) for each of the 4 classes, from top to bottom: bulgy, intermediate-$\eta$, pseudo-bulge, bulgeless. 
The solid lines correspond to the SBP of the global model, dashed lines correspond to the disk component and dotted lines to the bulge component. 
Right: SDSS $gri$ color-composite cutout image of each source considered. Image size is fixed (to $\sim 60\arcsec$ in side). }
\end{figure*}

The final step of this stage was to identify pseudo-bulges among the group of 42 galaxies that revealed an SBP where two structural components were obvious and compelling (a disk plus a bulge with S\'ersic index $\eta_b$ assuming various values).
To do so, a frontier value for $\eta_b$ was sought. In general terms, and since $\eta$ varies continuously from a Gaussian profile (0.5) through an
exponential disk (1.0) and to a classical bulge ($\sim$4), it is not
easy to objectively establish a boundary value separating bulgy galaxies
from those hosting a pseudo-bulge, or bulgeless ones. After a careful visual
inspection of the images and the results from GALFIT, and by
considering previous studies \citep{bell08,gadotti09,barentine12}, we finally adopted the conservative range of $\eta_b<1.5$ to identify pseudo-bulges - these total 29 in our sample. The remaining 13 galaxies will be henceforth referred to as intermediate-$\eta$ bulge galaxies. To summarize this procedure and the adopted classes, we present in table \ref{tab_classes}
the selection criteria applied to SBPs and the final number of galaxies in each class for our 106 objects.
Typical examples of galaxies of each class may be found in figure \ref{fig_galaxyprofiles}, and respective redshift distributions are given in figure \ref{fig_historedshift}. 
This figure also shows that there is no trend on the frequency of bulge type with redshift, so we expect that any change there may be on the physical resolution of SDSS images along the redshift interval probed by our sample will have no major impact on the fits to the galaxies SBPs
The reader is referred to \cite{coelho13} for more details on the fitting procedure and construction of the 1D surface brightness profiles shown in figure \ref{fig_galaxyprofiles} for illustration purposes. 

\begin{figure*}
\centering
\includegraphics[width=\linewidth]{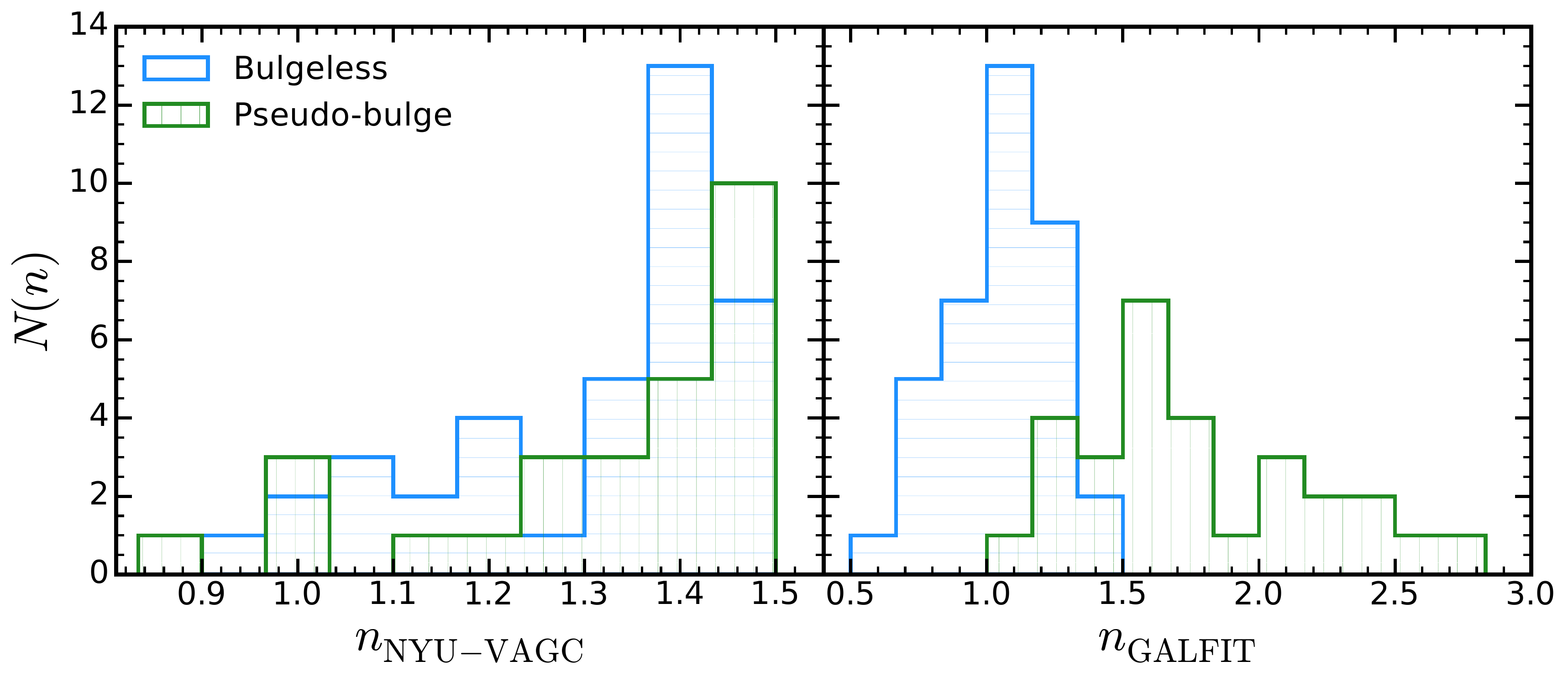}
\caption{\label{fig_histosersic}
Results of the test comparing the $\eta$ values issued from fitting a single S\'ersic model in 2D (with GALFIT; on the right) and in 1D (by \citealt{blanton05} - values available in the NYU-VAGC catalog; on the left). The galaxies used for this test are the ones we classified (by a different procedure - see text and equation \ref{multisersic}) as bulgeless (in blue) and pseudo-bulge galaxies (in green).}
\end{figure*}

A concluding verification was carried out on the performance of GALFIT as compared to the specific
adaptation of 1D surface photometry by \citet{blanton05}. After defining our classes, we fitted the SBPs of our bulgeless and pseudo-bulge galaxies with a single S\'ersic model (equation \ref{sersic}), and compared our values
for $\eta$ with those reported in the NYU-VAGC catalog for the same r-band images. Fig. \ref{fig_histosersic} reports the results, where we find that our 2D analysis yields a better
separation between bulgeless and pseudo-bulge
galaxies.  An additional reason for this may also lie in the automatized treatment
of the whole SDSS dataset by \citet{blanton05}, that obviously cannot handle the
particularities of individual galaxies, which is perfectly acceptable in
million-object catalogs used for large statistical studies.

\begin{table*}
\caption{Structural classes of galaxies considered, definition of each class
  based on the exponent $\eta$ of the best-fitting 2D S\'ersic model and
  number of objects in each class. Profile decomposition was made
  using a combination of two S\'ersic laws (see equation \ref{multisersic}): one with fixed $\eta_d=1$ (to
  model the disk) and another one with the S\'ersic index $\eta_b$ let free 
  when fitting the bulge component. Pure-disk galaxies obviously do not require an additional
  S\'ersic component whereas bulgy galaxies were fitted by a single S\'ersic model, without need 
  for an additional underlying disk.}
\label{tab_classes}
\centering
\begin{tabular}{ccc}
\hline
Structural class & interval of S\'ersic index for objects in this class & number of galaxies \\
\hline
bulgeless galaxies & $\eta_d=1$, no bulge component & 38 \\
pseudo-bulge galaxies & $\eta_d=1$, $\eta_b<1.5$  & 29 \\
intermediate-$\eta$ bulge galaxies & $\eta_d=1$, $1.5 < \eta_b < 3.5$   & 13 \\
bulgy galaxies & $\eta_b > 3$, no disk component   & 26 \\
\hline
\end{tabular}
\end{table*}

\vspace{2cm}

\section{Spectral synthesis of stellar populations}\label{sec:sps}

The analysis of the SDSS spectra of our 106 galaxies of different
structural classes was undertaken using the stellar population
synthesis code {\sc starlight}
\footnotemark{}\footnotetext{http://www.starlight.ufsc.br/}
\citep{fernandes05} to model the spectral energy distribution (SED) of
galaxies with a linear superposition of $N_\star$ Simple Stellar
Populations (SSPs). The best-fitting solution from {\sc starlight} is
obtained via standard $\chi^2$ minimization with a non-uniform
sampling of the parameter space. This code employs advanced
statistical mechanics techniques, such as simulated annealing and
multiple independent Markov Monte Carlo Chains to avoid convergence 
into local minima. 
Additionally, the \citet{gelru92} convergence criterion is implemented 
for a better determination of the global minimum solution.

The star formation and chemical enrichment history of a galaxy is obtained
from the best-fitting light population vector $\vec{x} =
(x_1,...,x_{N_\star})$, i.e. the set of fractional contributions of
stellar populations with different ages and metallicities. 
The mass population vector $\vec{\mu}$ is derived indirectly, after combining
$\vec{x}$ with the corresponding SSP mass-to-light ratios
(M$_\star/$L). In order to have a more realistic modeling, extinction
and kinematic parameters are also taken into account. 
For the former a uniform dust screen and the \citet{cardelli89} reddening law
were adopted and for the latter we use a Gaussian kernel $G(v_\star, \sigma_\star)$
that is determined by two parameters: the systemic velocity $v_\star$ and the
velocity dispersion $\sigma_\star$ that model the line shifts and
broadening effects, respectively.

Besides the main products issued from the {\sc starlight} fits, we can
also use these fits to accurately extract and measure any remaining
emission-lines. For this, we have subtracted the best-fitting stellar
SED from the observed spectrum in order to isolate the pure
emission-line spectrum. This procedure is crucial, especially for
bulgy galaxies that usually exhibit very weak emission-lines, but also
possibly for several of the remaining objects (due to our selection
described in section \ref{sec_sample}).

\subsection{Inferred physical properties from multiple {\sc starlight} fits}

In our analysis, we make use of the evolutionary synthesis models from
\cite{bruzual03} to compute a set of 150 SSPs that comprise 25 ages
(from 1 Myr to 15 Gyr, covering all evolutionary phases of single
bursts)
for 6 metallicities: 1/200, 1/50, 1/5, 2/5, 1, 2.5 $Z_\odot$,
where the solar value $Z_\odot$ is 0.02. These models are based on
the ``Padova 1994'' evolutionary tracks
\citep{alongi93,bressan93,fag94a,fag94b,girardi96} and the
\citet{chab03} initial mass function between 0.1 and 100 M$_{\odot}$.


Due to the probabilistic nature of the solutions from {\sc starlight}, 
the population vector obtained for a spectrum can slightly vary among different 
fits. In order to estimate the impact of this inherent dispersion in model fits 
on the spectral synthesis parameters, we have devised a new approach where 
50 {\sc starlight} runs were done for each galaxy spectrum to better quantify 
systematic uncertainties in the mean stellar age and metallicity, stellar mass and velocity
dispersion, among other spectral synthesis quantities, and also to
have a better error statistics on the emission-lines and line ratios
measured after the removal of the underlying stellar contribution.

We note that all results obtained with {\sc starlight} apply to the
SDSS fiber spectra (aperture $3\arcsec$), which enclose only the
central parts of galaxies, so the spectral synthesis parameters refer
to a projected area 
with a diameter between $\sim$1.2 and $\sim$3.4 kpc 
for the most nearby and most distant sources in our galaxy sample,
respectively.

\section{Pseudo-bulge galaxies in our sample - similar to disks or rather to classical bulges?}\label{sec_results}

In order to gain insight into the nature and most likely origin of
pseudo-bulges we compared several parameters obtained both from the
structural analysis and stellar population synthesis,
previously described in sections \ref{sec_galfit} and \ref{sec:sps},
respectively.
As stated before, the aim is to verify whether there is evidence for distinguishing 
pseudo-bulges from classical bulges, and lending support to a disparate nature and 
evolutionary scenario between both.

\subsection{Structural parameters}\label{sec_structure}

\begin{figure}
\centering
\includegraphics[width=\linewidth]{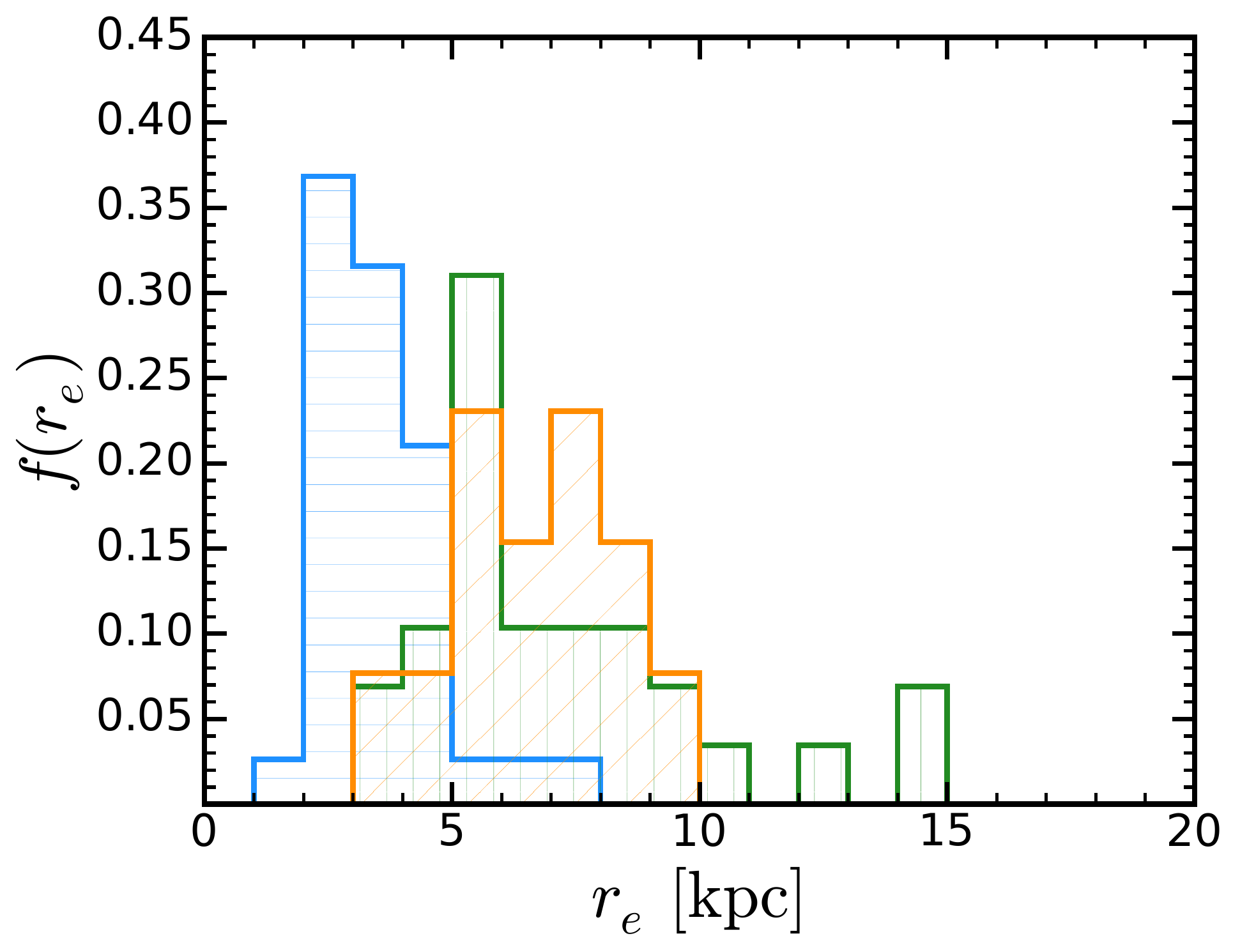}
\caption{Normalized histograms of the values of the disk effective radius,
  $r_{e,d}$, for the three classes having a disk component: blue
  is for pure disks, green refers to
  pseudo-bulge galaxies, while the orange histogram shows
  values for the intermediate-$\eta$ bulge galaxies.}
\label{fig_redhisto}
\end{figure}

\begin{figure*}
\includegraphics[width=\linewidth]{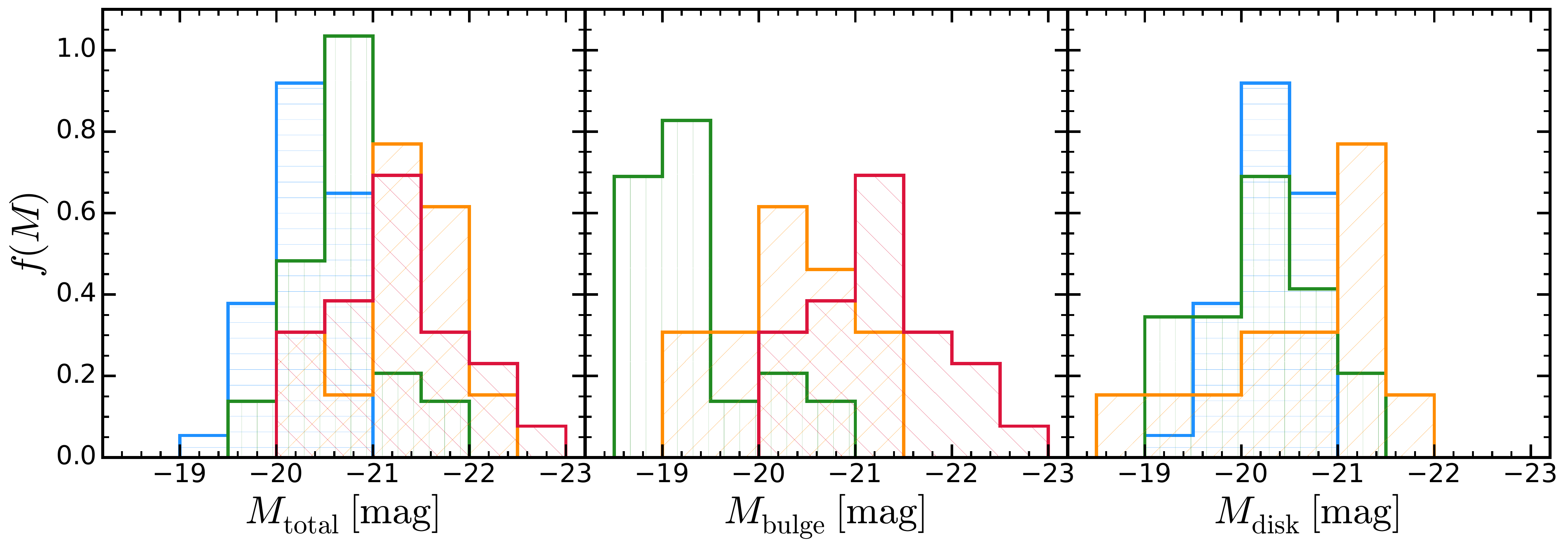}
\caption{\label{fig_maghistos} Normalized histograms of the total
  (left panel), bulge (middle panel) and disk (right panel) absolute
  magnitudes in the r-band for all structural classes: blue is
  for pure disks, green refers to pseudo-bulge galaxies,
  the orange histogram shows values for the intermediate-$\eta$
  bulge galaxies and the red is for bulgy
  galaxies.}
\end{figure*}

Table \ref{tab_structure} and figures \ref{fig_redhisto} and
\ref{fig_maghistos} summarize the trends that our analysis 
of the structural parameters reveal.

Although the dispersion of values is somewhat large, along the sequence  
bulgeless $\rightarrow$ pseudo-bulge $\rightarrow$ intermediate-$\eta$
bulge galaxies, there is a monotonous increase of the effective radius 
of the disk component, $r_{e,d}$ ($\approx$1.7 exponential scale lengths), 
for galaxies with similar absolute disk magnitudes in the r band. 
This is apparent from the 
histogram in Fig.~\ref{fig_redhisto}, 
as well as in the median value (Table \ref{tab_structure}; mean values 
reflecting the same trend). 
A similar though less pronounced behavior was obtained by
\citet{gadotti09} for a sample of nearly 1000 SDSS galaxies selected
and analysed in a different way.

As for absolute magnitudes (Fig. \ref{fig_maghistos} and Table
\ref{tab_structure}), while disks and pseudo-bulge galaxies span quite
the same range of both total and disk luminosities, intermediate-$\eta$
bulge galaxies tend to have slightly brighter disks on average. 
The two classes hosting classical bulges (bulgy and intermediate-$\eta$) 
reach up to brighter absolute total magnitudes, likely due to the relevant
bulge contribution.

\begin{table*}
\caption{Median and dispersion (standard deviation) values for some of
  the parameters inferred from the structural analysis for each class of
  galaxies (presented in section \ref{sec_galfit}): 
  total absolute r-band magnitude ($M_r (\mathrm{total})$), bulge and disk magnitudes ($M_r (\mathrm{bulge})$ and $M_r (\mathrm{disk})$, respectively)
  and effective radius $r_{e,d}$ of the disk component in kiloparsec.}
\label{tab_structure}
\centering
\begin{tabular}{ccccc}
\hline
Structural class & $M_r (\mathrm{total})$ & $M_r (\mathrm{bulge})$ & $M_r (\mathrm{disk})$ & $r_{e,d}\ [\mathrm{kpc}]$ \\
\hline
bulgeless galaxies & $-20.3 \pm 0.6$  & -- & $-20.3 \pm 0.6$ & $3.4\pm1.2$ \\
pseudo-bulge galaxies & $-20.7 \pm 0.5$ & $-19.3 \pm 0.6$ & $-20.3 \pm 0.6$ & $6.1\pm2.9$ \\
intermediate-n bulge galaxies & $-21.3 \pm 0.6$ & $-20.2 \pm 0.6$ & $-20.7 \pm 0.8$ & $6.8\pm1.6$ \\
bulgy galaxies & $-21.1 \pm 0.6$ &  $-21.2 \pm 0.6$ & --  & -- \\
\hline
\end{tabular}
\end{table*}

By comparing galaxies having a disk component in
their SBP (i.e. a non-null first term in equation \ref{multisersic}),
the trends shown in this section suggest that the exponential disks of
pseudo-bulge galaxies are more diffuse (larger values for $r_{e,d}$) than those of bulgeless galaxies, since they share quite comparable 
absolute magnitudes.  
One should bear in mind though that this might
be partly due to the addition of a second component to the fit for the
first class of galaxies: the S\'ersic function that models the bulge
component (mainly the central light excess), leaves the pure
exponential to model best the outer, fainter regions of the galaxy to
a larger extent when compared to the single disk fit applied to
bulgeless galaxies. Despite the fact that this may artificially
increase the size of the disk by the simple additional presence of a
central component, the difference in sizes observed in Figure
\ref{fig_redhisto} cannot be 
attributed to the fitting procedure.
This was tested by applying the 2-component fit (equation \ref{multisersic}) to bulgeless galaxies - in these conditions,
the effective radii of their disk component remained systematically
smaller than for pseudo-bulge galaxies, fitted in exactly the same
way.

Behind the trends observed in Figure \ref{fig_maghistos} (left panel) is likely a well known correlation:  classical bulges are usually found in galaxies more massive (and thus brighter) than those hosting pseudo-bulges and these, in turn, are more massive (and brighter) than pure disk galaxies \citep[e.g.][]{fisher11}. We verify this trend in our sample since the   estimate we perform - through equation (1) of \citet{bell08} - gives mean values of 10.2, 10.3 and 10.5 for the logarithm of the total stellar mass of, respectively, disk galaxies, galaxies hosting pseudo-bulges and the remaining galaxies (hosting classical bulges). A wider discussion on the impact of the mass of galaxies on the properties of bulges is, however, beyond the scope of this paper.

\subsection{Stellar population synthesis parameters}\label{sec:synth}

\begin{figure*}
\includegraphics[width=0.49\textwidth,trim=0pt 10pt 0pt 90pt, clip]{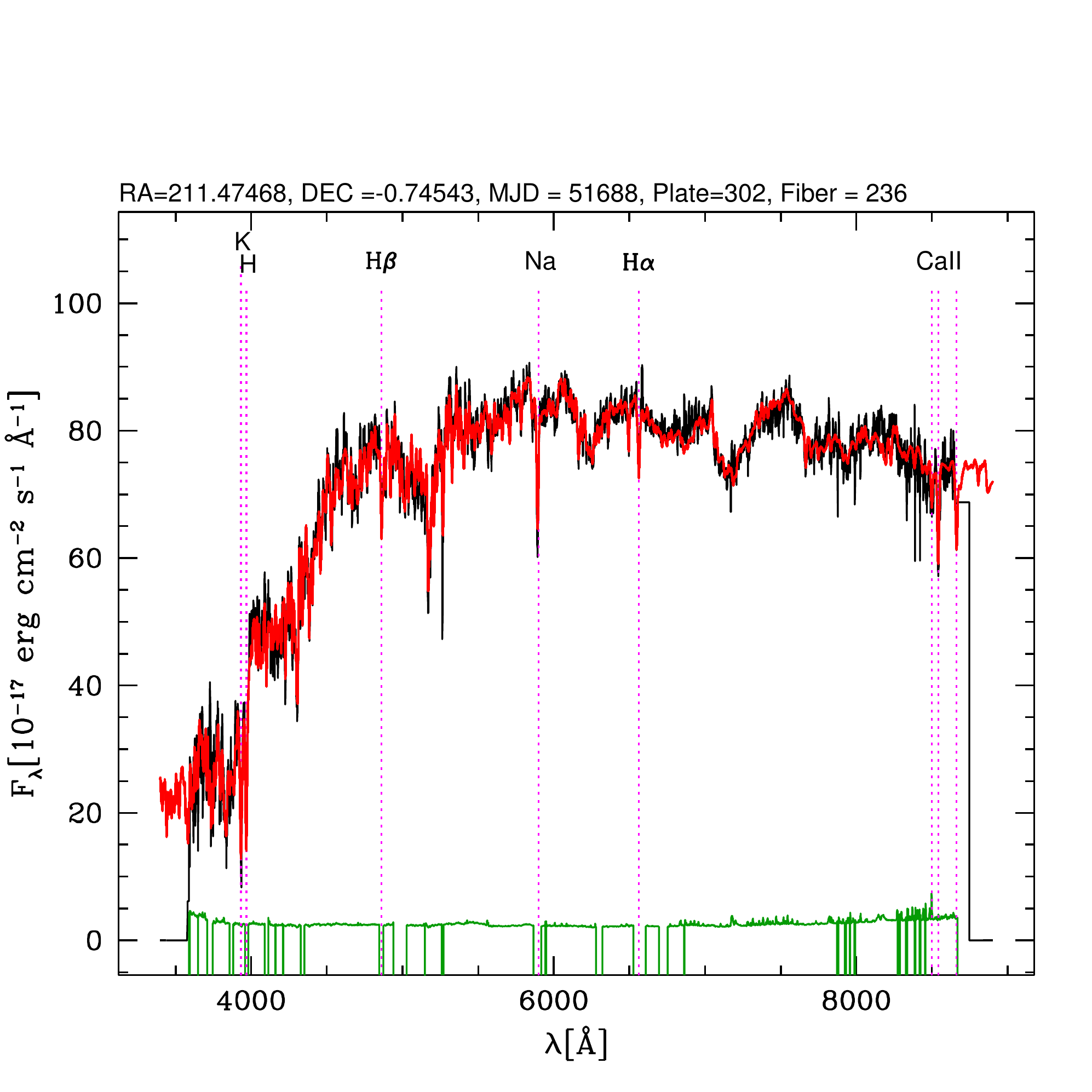}
\includegraphics[width=0.49\textwidth,trim=0pt 10pt 0pt 90pt, clip]{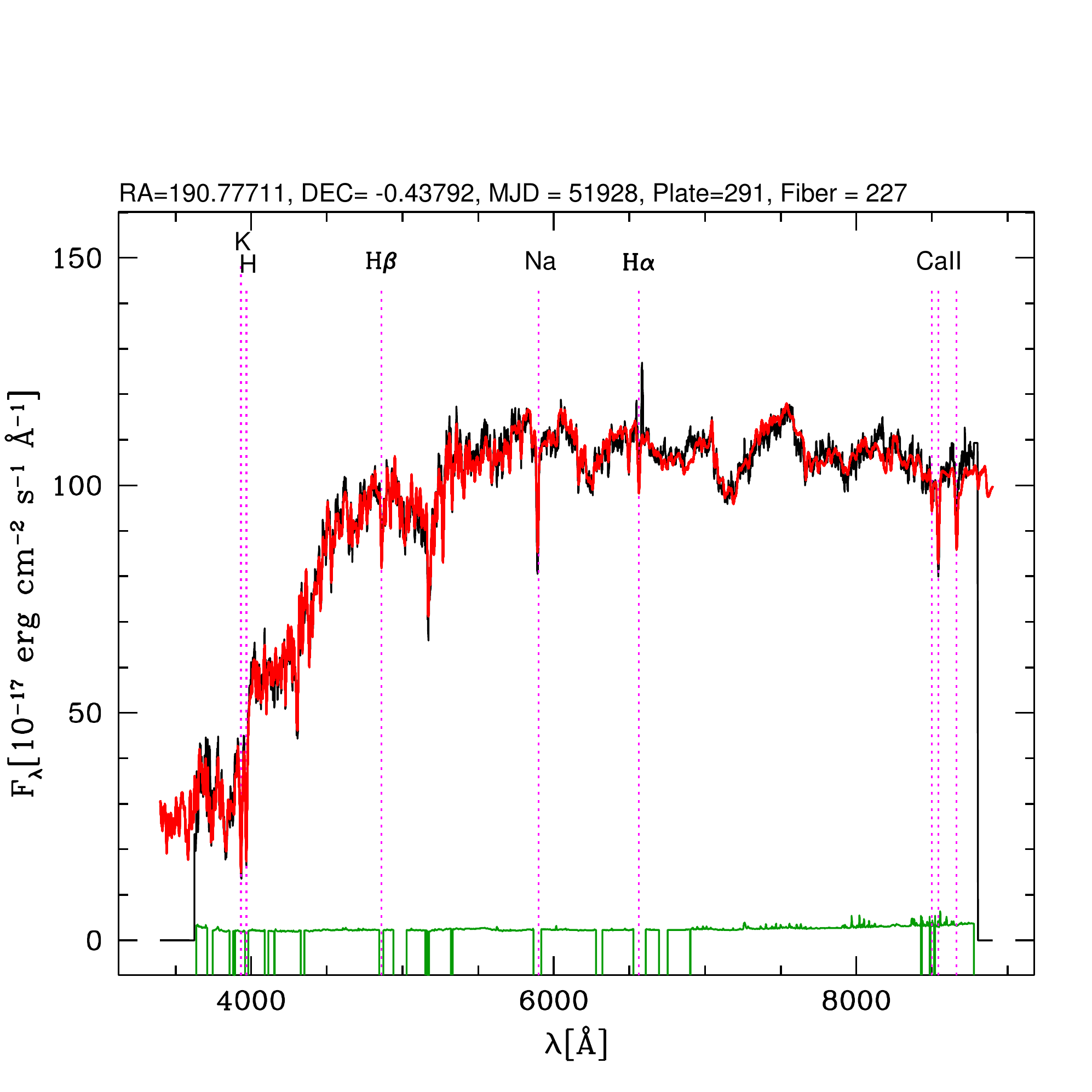}
\includegraphics[width=0.49\textwidth,trim=0pt 10pt 0pt 90pt, clip]{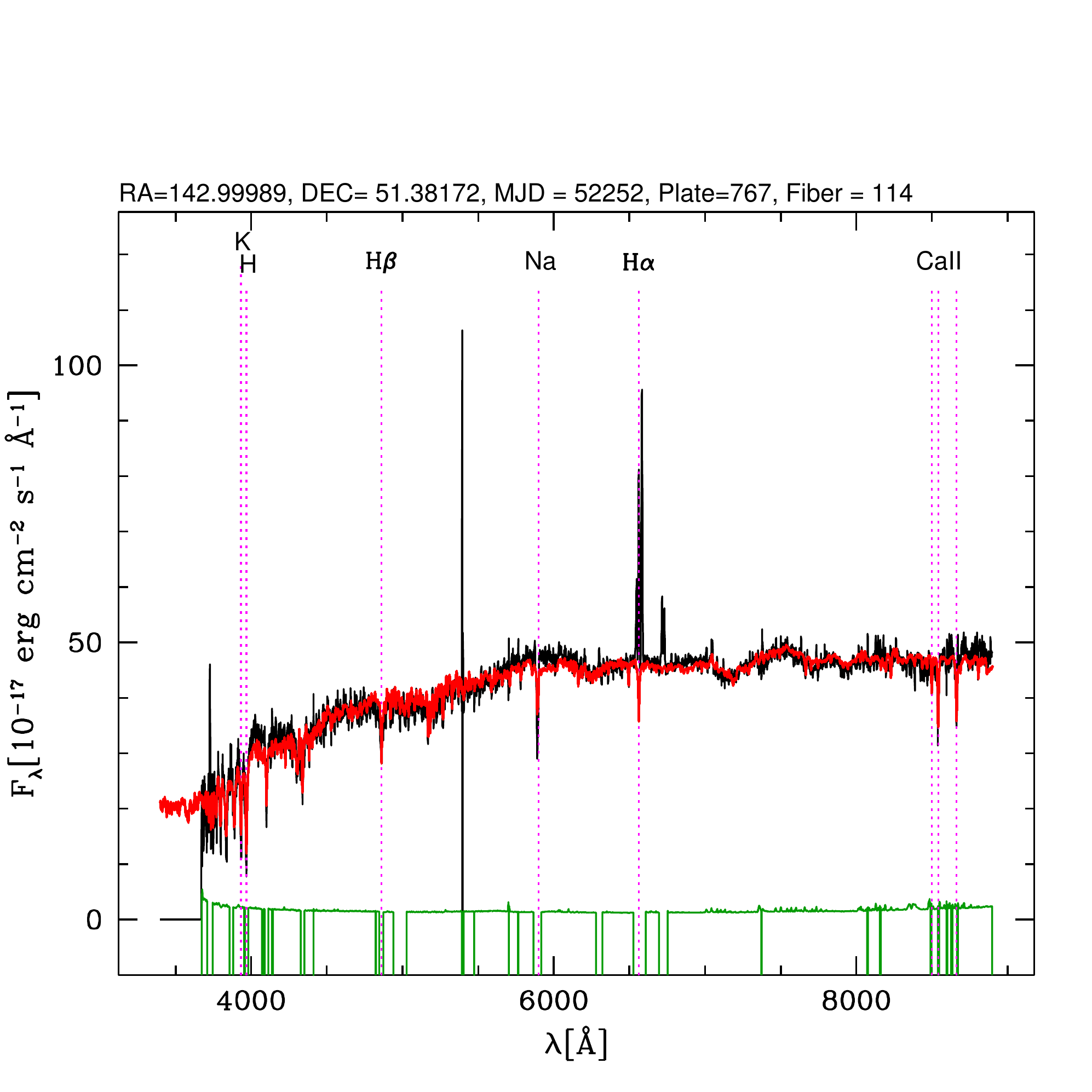}
\includegraphics[width=0.49\textwidth,trim=0pt 10pt 0pt 90pt, clip]{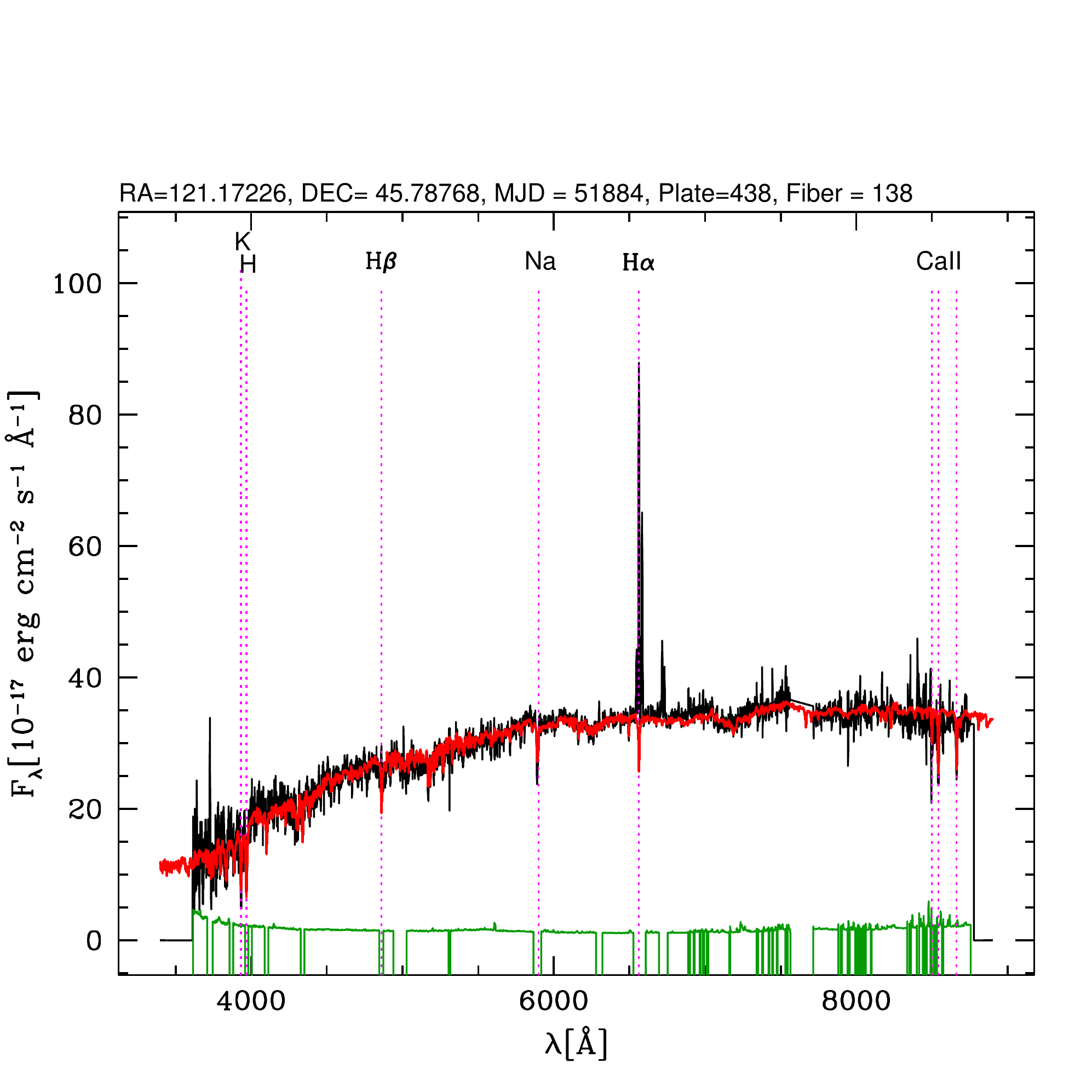}
\caption{SDSS spectra in rest-frame and corrected for Galactic extinction using the dust maps from \citet{schlegel1998} with additional corrections described in \citet{schlafly2011}. In each panel, the observed spectrum is in black, the red curve combines the 50 fits from \textsc{STARLIGHT} and green is the error spectrum (where negative values indicate wavelengths masked prior to the fit due to spurious pixels and/or emission lines). Each of the 4 classes is represented here: bulgy (top left), intermediate (top right), pseudo-bulge (bottom left) and bulgeless (bottom right).}
\label{fig_specs}
\end{figure*}

Typical spectra of objects assigned to our structural classes are
shown in figure \ref{fig_specs}. Different spectrophotometric
characteristics are evident: while classical bulges tend to be redder
and essentially lack conspicuous emission-lines, pseudo-bulges and the
central regions of bulgeless galaxies are bluer and show weak
emission-lines, pointing towards recent or ongoing star-formation.

Running {\sc starlight} 50 times for each galaxy spectrum allows us to
estimate the uncertainties associated with various quantities of
interest. We also expect this procedure to 
offer
a more reliable
statistical approach, as already stated in section \ref{sec:sps}.
Note that we have considered all SDSS spectra without imposing any
restriction in terms of signal-to-noise nor any other quality
criterion. In any case, the average S/N in the continuum region
(4730--4780 \AA) is $22.6 \pm 6.9$ for the whole sample, the minimum
S/N value being 9.3 (for 1 galaxy only),
quite sufficient for a good fit.

The population synthesis models yielded very similar trends for bulgy and intermediate-$\eta$ bulges (i.e. what we have been calling classical bulges or henceforth refer to simply as bulgy galaxies)
with a large overlap
in what regards many important properties and parameters.
For this reason we 
merged 
these two classes from here onwards - simply called
\emph{bulgy}, standing for bulge dominated galaxies - since they are
consistently alike in their stellar populations.
 
In this section (and in the following one), all plots show both the individual {\sc starlight}
realizations for each galaxy (as small dots) as well as the corresponding median value (large circles). Bulgy, pseudo-bulge and bulgeless galaxies are distinguishable through colors, being coded as red, green
and blue, respectively. 
A median value of the main parameters obtained from {\sc starlight} and discussed
next are given in Table \ref{tab_parameters} for each galaxy class, along with the respective 
1$\sigma$ dispersion.

\begin{figure}
\includegraphics[width=\linewidth]{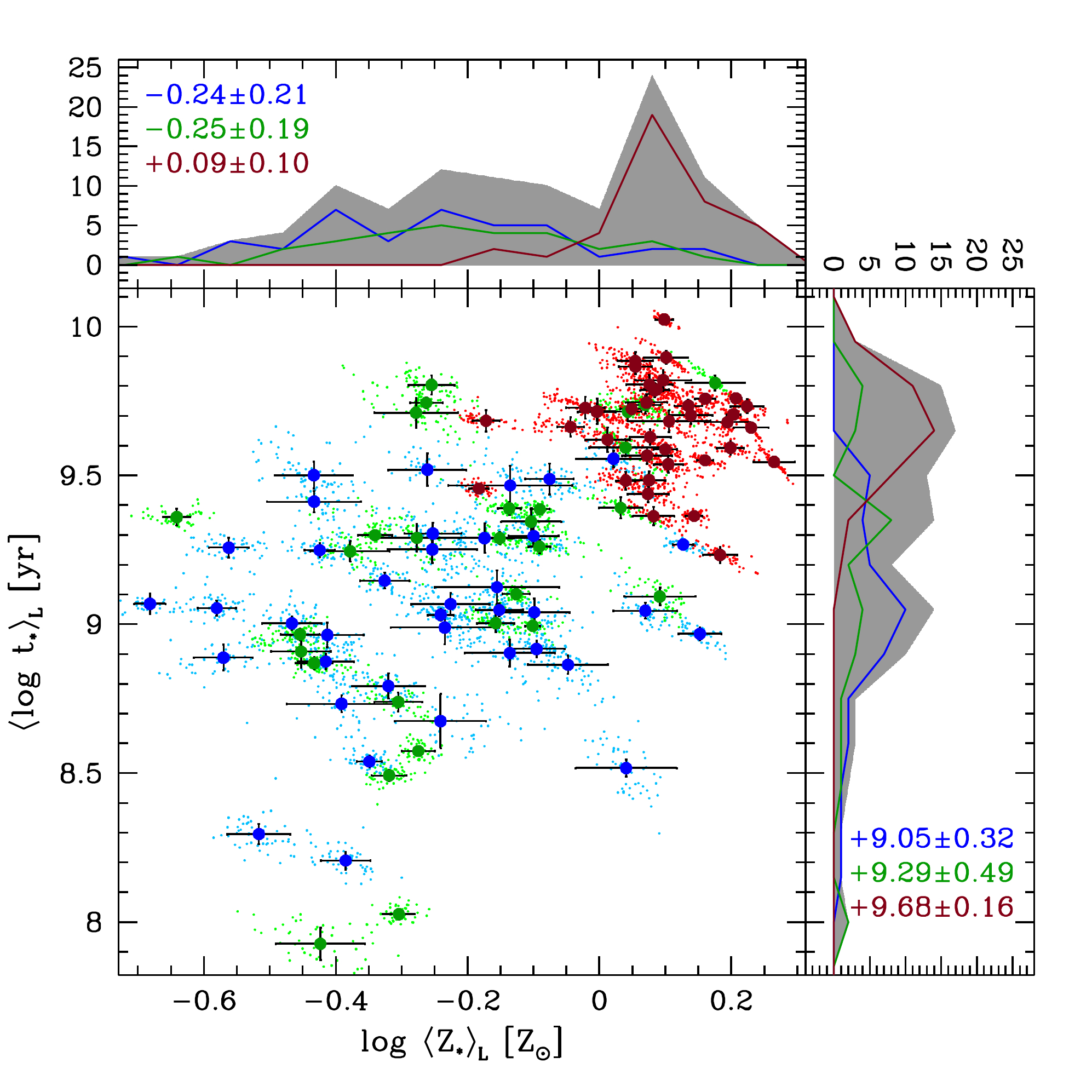}
\caption[]{Mean stellar metallicity $\langle Z_\star \rangle_L$ (in
  solar units) versus mean stellar age $\langle \log t_\star
  \rangle_L$, both luminosity-weighted, for bulgeless (blue),
  pseudo-bulge (green) and bulgy (red) galaxies (we note this class henceforth merges all $\eta_b > 1.5$ galaxies, irrespective of having a disk or not). For each galaxy, the 
  50 {\sc starlight} individual runs are shown as small dots, while
  the large filled circle corresponds to the median value with 1$\sigma$
  (standard deviation) error bars. Histograms are also shown for each
  class (same color coding): on the top panel for $\langle Z_\star
  \rangle_L$, and on the right-hand-side panel for $\langle \log
  t_\star \rangle_L$; the grey shaded area depicts the total
  distribution of our galaxy sample. The small colored numbers in each
  histogram panel are the median values, with 1$\sigma$, for each
  structural class (with the same respective color coding). The y-axis values represent the number of galaxies.}
\label{fig:t_LxZ_L}
\end{figure}

In figure \ref{fig:t_LxZ_L} we show two luminosity-weighted
quantities: the mean stellar metallicity $\langle Z_\star \rangle_L =
\sum_{j=1}^{N_\star}x_j Z_j$ as a function of the mean stellar age
$\langle \log t_\star \rangle_L = \sum_{j=1}^{N_\star} x_j \log t_j$,
where $Z_j$ and $t_j$ are the metallicity and age of the j$^{\rm th}$
SSP. Despite the spread in the data, especially apparent for later
types, we can see a conspicuous increase along the sequence from bulgeless,
pseudo-bulges to bulgy galaxies, where the median values for the  
stellar age were determined to be 1.12 $\rightarrow$ 1.95 $\rightarrow$ 4.79 Gyr,
while the mean stellar metallicity goes from 0.6 to 1.2 $Z_\odot$.  
This suggests that centers of bulgeless galaxies and pseudo-bulges span a larger
interval on both parameters but, on average, have younger and less
metallic stellar populations in comparison with classical bulges. 
The main reason might be due to the fact that the two former classes are typically 
more gas-rich and possibly sustain ongoing low-level star formation, 
whereas almost gas-devoid bulgy galaxies have ceased forming stars early-on
after depletion of their gas supply and simultaneous increase of their metal content
(see section \ref{subsec:MAH}).

\begin{figure}
\includegraphics[width=\linewidth]{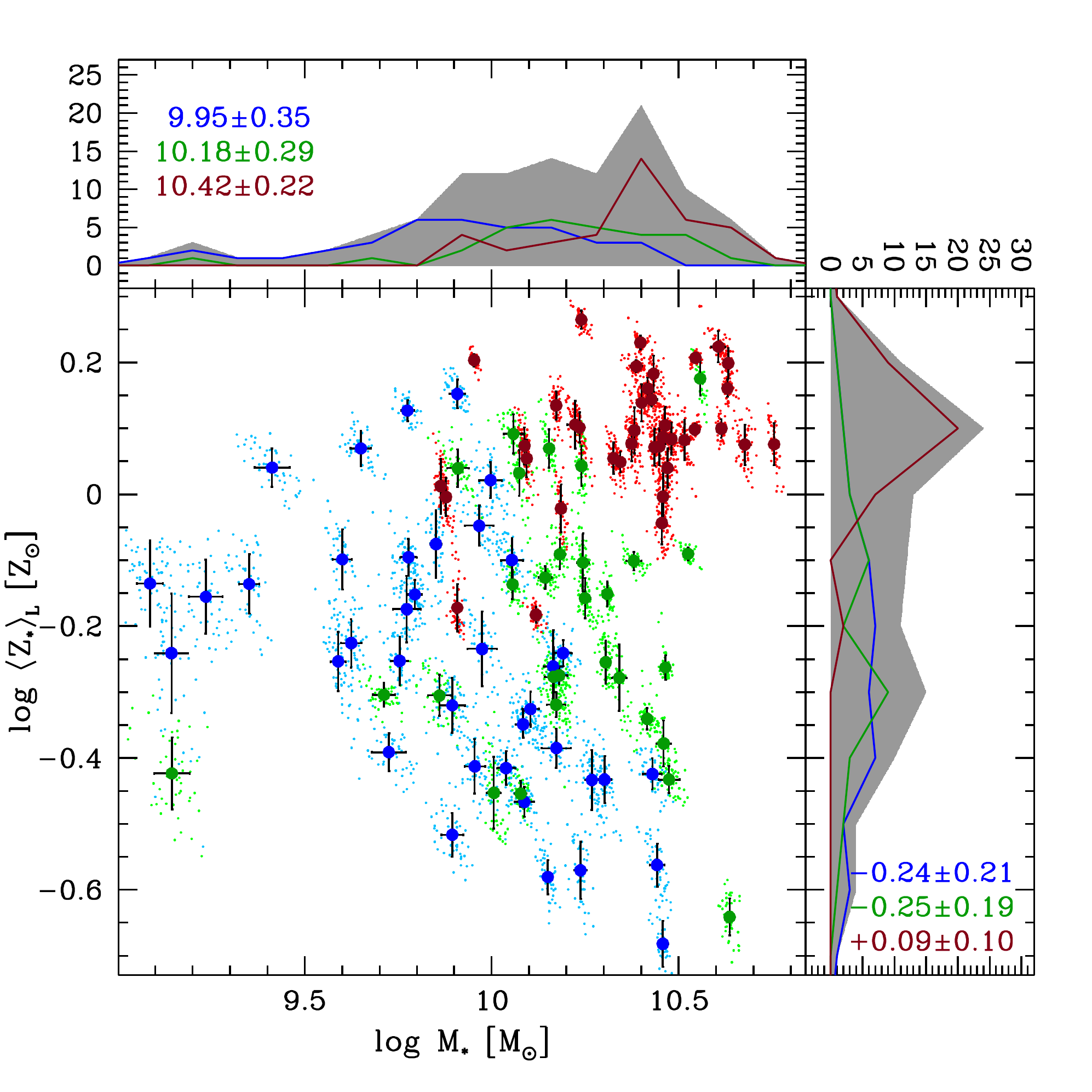}
\caption[]{Logarithm of the total presently available stellar mass
  (i.e. the stellar mass ever formed, after correction for the fraction
  of returned mass into the ISM) as a function of the light-weighted
  mean stellar metallicity. This diagram delineates a trend similar to
  the Tremonti et al. (2004) mass-metallicity relation, but with
  respect to the stellar values and only using our sample of red
  galaxies. Symbols, color coding and overall structure of this figure
  are as in Fig. \ref{fig:t_LxZ_L}.}
\label{fig:Ms_Zm}
\end{figure}

The mass-metallicity relation for our sample of galaxies is shown in
figure \ref{fig:Ms_Zm}, where the stellar mass has been
corrected for the returned fraction to the inter-stellar medium (ISM) according to
\citet{bruzual03}. Although the distributions largely overlap with regard to
the total stellar mass enclosed within the fiber, there is a slight
increase, on average, from bulgeless to bulgy galaxies (see Table
\ref{tab_parameters}), that likely follows the similar increase in the
bulge luminosity (see Table ~\ref{tab_structure}).

\subsection{Kinematics of the central regions}\label{subsec:FJ}

We have further investigated whether any of our structural classes deviates from 
the Faber-Jackson relation \citep[][hereafter FJ]{fj76} for classical bulges and early-type galaxies.
To this end, we adopted as reference the relation by \citet{labarbera10} who 
analyzed a sample of $\sim$40 000 nearby early-type galaxies by fitting
the relation $\log \sigma_\star = \lambda_0 + \lambda_1 \times
(\log L + 0.4X)$ for the grizYJHK photometric bands, where  
$\sigma_\star$ is the stellar velocity dispersion and $L$ the galaxy 
luminosity in the respective band. 
These authors have obtained the linear regression coefficients
$\lambda_o$ and $\lambda_1$ for the FJ relation after assuming a
magnitude limit $X$ in the corresponding wavebands. For the SDSS
r-band, the magnitude limit and coefficients they derived are
$X=-20.60$, $\lambda_0 = 2.151 \pm 0.008$ and $\lambda_1 = 0.192 \pm
0.018$. 

\begin{figure}
\includegraphics[width=\linewidth]{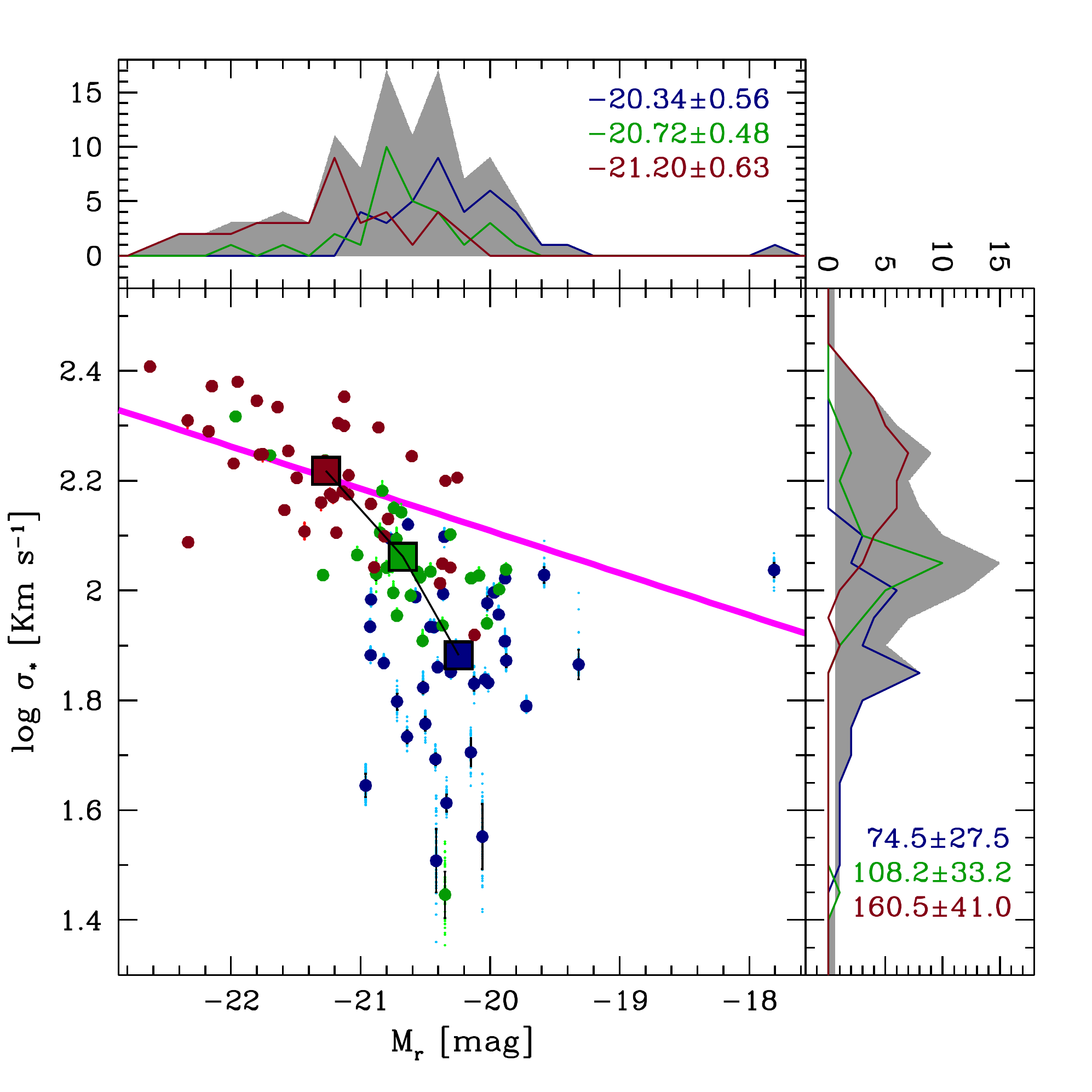}
\caption[]{Logarithm of the stellar velocity dispersion (in km
  s$^{-1}$) within the central region as a function of the total r-band magnitude (computed with
  GALFIT) for all galaxies belonging to the three classes now
  considered. Symbols and colors are as before
  (Figs. \ref{fig:t_LxZ_L}, \ref{fig:Ms_Zm}; but velocity dispersion
  values on the right panel are given directly in km/s), and we've now added
  large squares to mark, for each structural class, the median value
  of the parameter 
under study
(as in Table \ref{tab_parameters}). The thick magenta line shows the
  Faber-Jackson (FJ) relation from \citet{labarbera10} for early-type local
  galaxies. We can clearly see that bulgy galaxies fall within that
  relation, having a median value of -21.20 mag in the r-band and a mean 
  velocity dispersion of $\sim 161$ km/s. On the other hand, bulgeless
  and pseudo-bulges deviate strongly from the FJ relation.}
\label{fig:FJ}
\end{figure}

After plotting the logarithm of $\sigma_\star$ (obtained through {\sc
  starlight}) as a function of the total absolute magnitude for our
sample galaxies, we added the \citet{labarbera10} FJ relation (thick
magenta line in Fig. \ref{fig:FJ}). Bulgy galaxies seem to follow the
relation quite well whereas pseudo-bulges and disks are located
systematically below it, 
in agreement with the trends suggested by \citet{kk04} and also observed by e.g. \citet{zhao12} for a particular 
  sample of 75 bulges in isolated
Sb-Sc SDSS nearby spiral galaxies.  In our work, we are aware that we cannot
use $\sigma_\star$ as a ``clean'' measure of the velocity dispersion
in the central region of bulgeless and pseudo-bulge galaxies because
we expect these systems to show a significant degree of rotational
support, at variance with pressure-supported classical bulges.
Therefore, $\sigma_\star$ has to be seen as an upper-limit to the true
stellar velocity dispersion in these systems, which should only
reinforce the result we obtained. Finally, it is worth pointing out
that pseudo-bulges tend to lie in the intermediate region between pure
disks and bulgy galaxies, as evidenced by the median values of the
distributions - large squares in Fig. \ref{fig:FJ} and Table
\ref{tab_parameters}. Taken in combination with the previous results
on stellar ages, metallicity and stellar masses, one might infer that pseudo-bulges and the centers of
bulgeless galaxies possibly share a similar formation pathway, which
is distinct from the one driving the rapid growth and settlement of
classical bulges on the FJ relation.

\begin{table*}
\caption{Median and dispersion (standard deviation) values of several
  parameters obtained from fitting {\sc starlight} models to SDSS spectra that gather 
  the emission within the inner 3\arcsec\ of each sample galaxy. 
  Columns give, for each structural class now
  considered: mean stellar metallicity (in solar units); mean stellar
  age (in years); stellar mass (in solar units) corrected for the mass returned to 
  the ISM due to stellar winds and SNe explosions; V-band
  extinction; velocity dispersion of stellar absorption lines (in km/s);
  emission line ratios [NII]$\lambda$6583/H$\alpha$,
  [OIII]$\lambda$5007/H$\beta$ and H$\alpha$/H$\beta$.}
\label{tab_parameters}
\centering
\tabcolsep=0.15cm
\begin{tabular}{ccccccccc}
\hline
Structural class & log $\langle Z_\star \rangle_L$ & log $\langle \log t_\star
  \rangle_L$ & log M$_*$ & $A_V$ & $\sigma_\star$ & log [NII]/H$\alpha$ & log [OIII]/H$\beta$ & log H$\alpha$/H$\beta$     \\
\hline
bulgeless galaxies & $-0.24 \pm 0.21$ & $9.05 \pm 0.32$ & $9.95 \pm 0.35$ & $0.74 \pm 0.37$ & $74.5 \pm 27.5$ & $-0.36 \pm 0.10$ & $-0.38 \pm 0.32$ &  $0.79 \pm 0.13$           \\
pseudo-bulge galaxies & $-0.25 \pm 0.19$ & $9.29 \pm 0.49$ & $10.18 \pm 0.29$ & $0.84 \pm 0.39$ & $108.2 \pm 33.2$ & $-0.17 \pm 0.18$ & $-0.18 \pm 0.37$ &  $0.80 \pm 0.27$        \\
bulgy galaxies & $+0.09 \pm 0.10$ & $9.68 \pm 0.16$ & $10.42 \pm 0.22$ & $0.03 \pm 0.08$ & $160.5 \pm 41.0$ & $0.07 \pm 0.17$ & $0.13 \pm 0.17$ & $0.32 \pm 0.32$                \\
\hline
\end{tabular}
\end{table*}

\subsection{Emission-lines and line ratios}\label{subsec:EmLines}

As mentioned in section \ref{sec:sps}, emission-lines were measured
after subtraction of the best-fitting stellar continuum model. Note
that this procedure was repeated 50 times for each SDSS spectrum, due to
the multiple {\sc starlight} fits obtained. 
Quite importantly, not in all 50 trials were we able to measure the emission-lines, 
especially for bulgy galaxies where nebular emission is generally very faint.

Table \ref{tab_parameters} shows that $A_V$, the $V$-band extinction
determined from models with {\sc starlight}, is remarkably distinct 
for the different classes:
there seems to be a dichotomy, separating bulgeless/pseudo-bulges from
bulgy galaxies. The former also span a similar distribution with regard 
to the gas-phase extinction, inferred through the Balmer decrement 
H$\alpha$/H$\beta$, with typical values higher than 2.86 (the
theoretical value for these recombination lines in H{\sc ii}
regions for typical conditions) and being clearly separated from bulgy galaxies.
Both parameters ($A_V$ and Balmer decrement) are obviously
correlated and their values lead us to infer that the centers of pseudo bulges 
and disk galaxies are more dust-enriched than classical bulges.

\begin{figure}
\includegraphics[width=\linewidth]{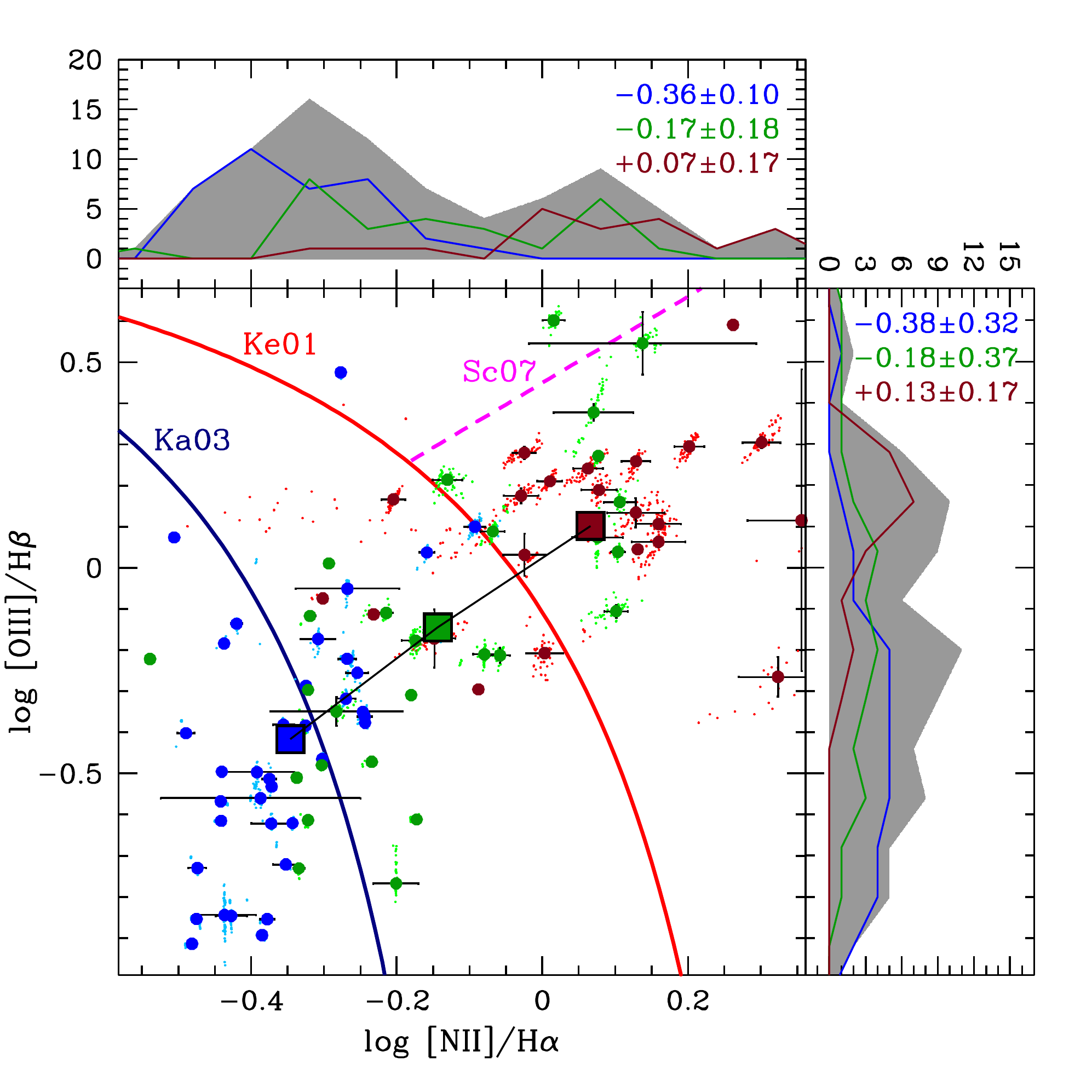}
\caption[]{BPT emission-line ratio diagnostic diagram for
  [NII]$\lambda$6583/H$\alpha$ versus
  [OIII]$\lambda$5007/H$\beta$. The demarcation lines in blue, red
  and magenta are the ones proposed by \citet[][Ka03]{ka03},
  \citet[][Ke01]{ke01} and \citet[][Sc07]{sc07}, respectively. They
  are used to classify galaxies according to the ionization source:
  Ke01 is a theorectical upper limit from stellar photoionization
  models, while Ka03 is used for demarcating star-forming galaxies; Sc07 is used
  to separate between Seyfert and LINER regimes. Symbols are as 
  before (see Fig. \ref{fig:FJ}).}
\label{fig:BPT}
\end{figure}

In order to further investigate the star formation activity of our sample
galaxies we have derived the classical \citet[][hereafter BPT]{bpt81}
diagram to classify optical emission-line spectra into
H{\sc ii}/Star-Forming (SF), Composites and LINER/Seyfert galaxies using the
[NII]$\lambda$6583/H$\alpha$ versus [OIII]$\lambda$5007/H$\beta$
emission-line ratios (see Fig. \ref{fig:BPT}). The adopted demarcation
lines are from \citet[][hereafter Ka03]{ka03} to select SF galaxies,
and the theoretical upper limit obtained from photoionization models
proposed by \citet[][hereafter Ke01]{ke01} to separate star-forming
from other ionization sources such as shocks, post-AGB stars and
AGNs. In order to discriminate Seyferts from LINERs, we 
adopted the
\citet[][hereafter Sc07]{sc07} 
demarcation line.

In terms of the ionization properties, we can clearly see a sequence
where star-forming galaxies tend to be bulgeless, pseudo-bulges are
mainly composites (though there is a large dispersion) and bulgy
commonly lie in the LINER regime. Since disky and pseudo-bulges are
found preferentially below the Ke01 demarcation line, we can conclude
that there is still ongoing star-formation activity, while extra
sources of ionization have to be invoked in bulgy systems to explain
their emission-line ratios, and only residual star-formation might be
in place. This is compatible with the hypothesis of disky/pseudo-bulges 
building up in the course of secular galaxy evolution, in contrast to  
bulgy galaxies that have formed the bulk of their stars much earlier and presently 
exhibit very low, if any at all, star-formation activity.  
It also shows that our selection minimizes
occurrence of star-forming objects in our sample \footnotemark{ }
 \footnotetext{As described in section \ref{sec_sample},
galaxy g-r SDSS colors are red: mean values range from 0.75 to 0.81 for the different classes of table \ref{tab_classes}, being rather typical of early-types \citep[e.g.][]{fuku95}. Individual galaxy colors are given in tables \ref{tab_allgals}-\ref{tab_intermediate}. Visual inspection of the images, also described in section \ref{sec_sample}, further concurred to eliminate objects with characteristics indicative of significant star formation.}
but does not exclude them completely, as expected.
Noteworthy is also that, despite being intrinsically red in terms of having 
a dominantly aged, red stellar population, some star formation is taking
place in these bulgeless and pseudo-bulge galaxies - enough to produce the 
gas excitation we observe.
These situations, however, are not a problem for this work
since the generally very faint nebular emission in these systems should not lead 
to any bias on our results, while rendering confidence to our morphological classification and SPS results.

\subsection{Mass assembly history of galaxies}\label{subsec:MAH}

\begin{figure}
\includegraphics[width=\linewidth, viewport=0 20 580 470]{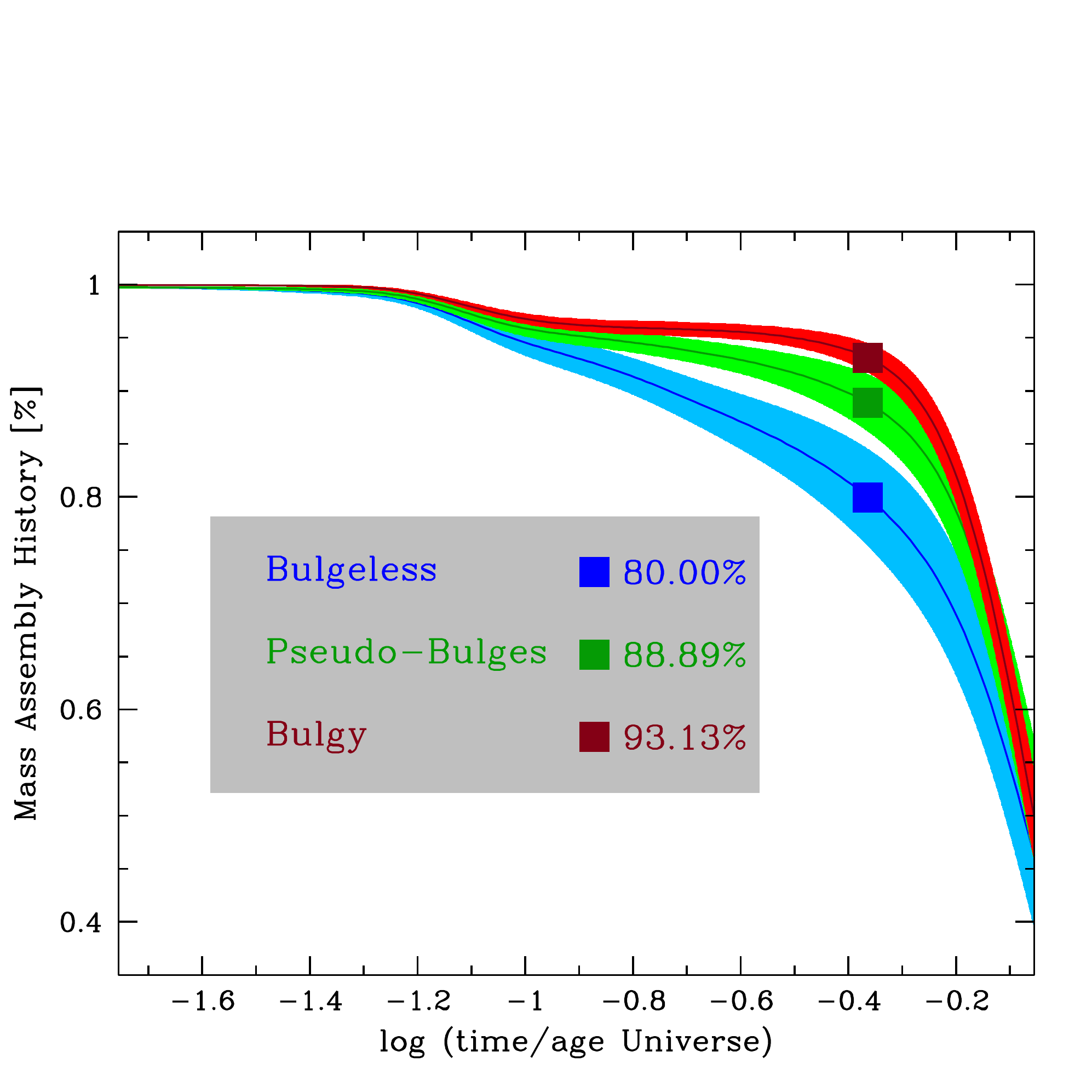}
\caption[]{Mass assembly history (MAH) as a function of lookback time,
  normalized to the age of the Universe, for our different structural
  classes: bulgeless (blue), pseudo-bulges (green) and bulgy galaxies
  (red). Each curve shows the median trend for each galaxy class under
  consideration, with the same color shading showing the dispersion
  of the values (including the multiple {\sc starlight} runs for each
  galaxy). Since this sample consists of optically red galaxies, the
  bulk of their stellar mass was formed more than 1 Gyr ago. However,
  the pace at which each class assembles its stellar mass is different, being slower in bulgeless, moderate in
  pseudo-bulges and faster in bulgy galaxies. The large filled squares
  mark the corresponding MAH values for an age of the Universe of
  $\sim$ 7.7 Gyr ($\sim$ 6 Gyr in lookback time) when bulgeless
  galaxies have assembled 80\% of their central stellar mass, while
  pseudo-bulge and bulgy galaxies have already formed $\sim 89\%$ and
  $\sim 93\%$ of their stellar mass (also in the inner $3\arcsec$),
  respectively.}
\label{fig:MAH} 
\end{figure}

An important insight that can be gained from spectral synthesis models
is the mass assembly history \citep[MAH,
  e.g.][]{heavens04,asari07,mcder15}, i.e. the cumulative stellar mass
fraction as a function of time, here determined within the central
$3\arcsec$ of the analyzed galaxies (Fig. \ref{fig:MAH}).  The curves
were obtained from the {\sc starlight} star formation histories of
galaxies and smoothed to 0.5 dex. The statistics include the 50
different {\sc starlight} runs for each galaxy in our sample and we
highlight the median trend for each structural class (bulgeless,
pseudo-bulge and bulgy). Since we have selected red galaxies from the
SDSS, we expect that the bulk of their stellar mass was formed rather
early-on for all distinct categories. Analyzing the curves in figure
\ref{fig:MAH}, we see that the assembly of the stellar mass within the
galaxy centers was accomplished more than 1 Gyr ago for our three
classes. However, the pace at which each structural class assembles
its central stellar mass is different, being slower in bulgeless,
moderate in pseudo-bulges and faster in bulgy galaxies. The large
filled squares mark the corresponding MAH values at an age of the
Universe of $\sim$ 7.7 Gyr where bulgeless galaxies have assembled
80\% of their inner stellar mass, while pseudo-bulge and bulgy
galaxies have already formed 89\% and 93\% of their fiber stellar
mass, respectively.

These results seem to be compatible with the secular evolution
scenario for pseudo-bulges, pointing to prolonged star forming activity that is fed by gas inflow
from the disk or the halo, leading to a gradual build-up of the (pseudo)bulge
at a slower pace. This star formation is a process still
ongoing, as demonstrated in the last section. It is worth reiterating that,
despite our optical color selection criteria, some of these
galaxies have emission lines reflecting ongoing star formation in a moderately 
dusty environment in their central regions.
This star formation seems to be relevant in light but probably not so much in the percentage of
central stellar mass formed: being enough to place them in the BPT
parameter space typical of H{\sc ii} regions and alike, and contributing to
secularly building-up the inner regions of these galaxies and the growth of the 
pseudo-bulge, it does not increase significantly the amount of stellar
mass already present - at least in the last Gyr.
 
On the other hand, classical bulges seem to assemble their 
stellar mass at a much faster rate, in a process consistent with a
violent galaxy merger or rapid gas collapse scenario. 
Additionally, they show no signs of significant recent star-formation activity in their centers, 
consist of old stellar populations only, 
and likely have almost no gas left to fuel star formation.
Our results are compatible with those obtained by \citealt{zhao12} through 
spectral fitting of bulges and pseudo-bulges in 75 isolated late-type galaxies located at lower redshift than our sample. 
This study concluded that either class has formed most of its stellar mass $\sim$10 Gyr 
ago, with the important difference of pseudo-bulges sustaining star-forming activity 
over a longer timescale than classical ones, in qualitative agreement with the secular 
evolution scenario.

\subsection{Discussion on possible aperture biases}\label{subsec_aperbias}

One should consider whether aperture effects 
might be inducing the trends unveiled by the population synthesis analysis. The fixed SDSS fiber aperture implies that, for higher redshift galaxies, the area probed by the spectra is physically larger (we recall it ranges from $\sim$1.2 to $\sim$3.4 kpc, in diameter, for our sample). We must then quantify, for galaxies having a disk+bulge structure, if their fiber spectra is contaminated by disk light. Fig. \ref{fig_historedshift} shows that the redshift distributions are quite similar, especially among galaxies having a bulge, so one would not expect the above hypothesis to have any impact on the observed tendencies. In any case, we have made some tests so as to be sure that our results concerning pseudo-bulges, and their apparent similarity with disks in several properties, are not driven by aperture effects.

\begin{figure}
\includegraphics[width=\linewidth]{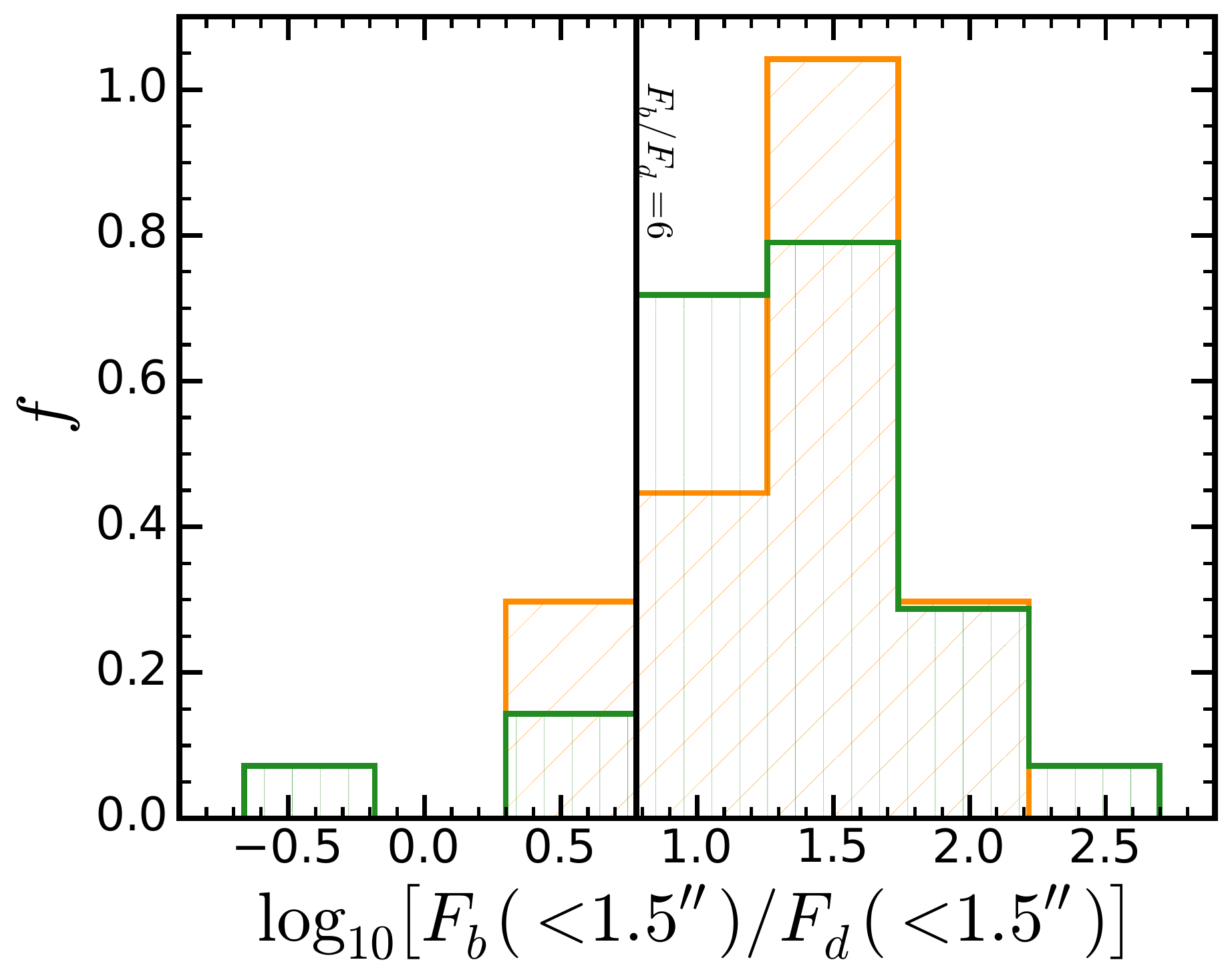}
\caption[]{Normalized histogram of the bulge-to-disk flux ratio inside the fiber aperture computed from the best fit models derived with GALFIT. The green histogram refers to
  pseudo-bulge galaxies, while the orange histogram shows
  values for the intermediate-$\eta$ bulge galaxies. The vertical black solid line refers to the limit for bulge dominance defined in \citet{zhao12}.}
\label{fig_apertures} 
\end{figure}

Following \citet[][see also references therein]{zhao12}, we computed the flux contribution provided by disk and bulge inside a region with radius $1.5\arcsec$ (respectively, $F_d(<1.5\arcsec)$ and $F_b(<1.5\arcsec)$), the size of SDSS fibers. These quantities are computed analytically from the integral of the S\'ersic function using the parameters obtained with GALFIT. Results for the ratio $r_{1.5} = F_b(<1.5\arcsec) / F_d(<1.5\arcsec)  \equiv L_b(<1.5\arcsec) / L_d(<1.5\arcsec)$ are shown in Fig. \ref{fig_apertures} for the two classes having a disk+bulge structure (i.e. pseudo-bulge galaxies and intermediate-$\eta$ ones). In the figure, we also mark the limit where the light from the bulge is about 6 times more than that from the disk within the $1.5\arcsec$ radius aperture; below this line lie objects where the disk contribution becomes important for the light collected by the SDSS fiber \citep{zhao12}. As most of our sample is above this limit, this clearly hints to aperture effects not being an issue in our sample: we thus expect that the parameters inferred from SPS for pseudo-bulges are not contaminated by the light of the underlying disk of the host galaxy.  

\section{Summary and Conclusions}\label{sec_conclu}
This article presents a comparative study of bulgeless, pseudo-bulge and 
\emph{bulgy} (classical-bulge and early-type) galaxies with respect to the star 
formation history, stellar velocity dispersion, metallicity and gas-excitation mechanisms 
in their central part. Our main goal is to obtain further observational constraints on the
question of whether pseudo-bulges are distinct from classical bulges in terms of their 
assembly history, as expected by the secular growth scenario for the former. 

Our sample was deliberately selected such as to comprise undistorted low-inclination
galaxies with spectrophotometric properties indicative of negligible levels of both ongoing 
star-forming activity and intrinsic obscuration by dust. 
This ensures 
a reliable morphological analysis and classification, and that the spectral fitting 
analysis of SDSS spectra is unbiased, 
thereby allowing for a robust comparison of the physical and evolutionary characteristics of 
the above classes. The latter were defined on the basis of the structural properties of our 
sample galaxies, as obtained through 2D bulge/disk image decomposition with GALFIT, and by 
demanding a minimum S\'ersic index $\eta > 1.5$ for classical bulges.
The redshift range of our sample ensures that the region probed by the SDSS fiber is adequate for sampling the stellar populations of the bulge without significant contamination from the underlying disk, which is an improvement on previous studies \citep{zhao12}.
Our sample (106 galaxies) comprises nearly equal parts of bulgeless (38), pseudo-bulge (29) and
bulgy (39) systems, in an attempt to increase the number of objects under study relatively to previous works related with this topic \citep[eg][]{moorthy06,macarthur09,zhao12}. Furthermore, a cut in stellar mass aims at avoiding introducing any strong dependence on this parameter by excluding low-mass galaxies.

The main results from this study may be summarized as follows:

\begin{itemize}

\item Our photometric analysis suggests that disks underlying pseudo-bulges typically have larger exponential scale lengths 
than bulgeless galaxies, despite similar integral disk luminosities. 
One can try to 
elaborate on whether this is related to the build up of the pseudo-bulge mass. 
Models of pseudo-bulge growth via satellite accretion \citep[such as the ones explored by e.g.][]{eliche06} 
show that these events lead to an increase of the disk scale length that depends on the mass of the satellite. 
This occurs due to the outward transport of disk material in the outer regions, combined with inward transport 
to the bulge in the inner regions. However, the very same physical argument (outward transfer of angular momentum) 
is also used to justify the expansion of galaxy disks and accompanying concentration in the inner regions, leading 
to bulge growth, but in the framework of secular evolution (see e.g. \citealt{kk04} and also \citealt{sachdeva15} 
and references to models therein).  
The results from our study {\em per se} do not permit discrimination between these two scenarios.

\item Spectral synthesis models of the stellar emission within the SDSS fiber aperture 
reveal a clear segregation of bulgeless and pseudo-bulge galaxies from bulgy ones 
with respect to the luminosity-weighted age ($t_\star$) vs metallicity ($Z_\star$):
Notwithstanding a large dispersion 
among galaxies within a class,
the former two classes were found to significantly differ (median of $<$2 Gyr and 0.6 $Z_\odot$, respectively) 
from classical bulges ($\sim$5 Gyr and 1.2 $Z_\odot$, respectively).
This contrasts with previous results: \citet{moorthy06,macarthur09, zhao12} find that pseudo and classical bulges do not have significantly different stellar populations. However, samples are selected in a noticeably different way and some of these works have very few objects, whereas a large scatter is present also in our case.

\item On the stellar mass ($M_\star$) vs $Z_\star$ plane, classical bulges predominantly populate 
the high-mass, high-metallicity locus ($M_\star \ga 10^{10}$ $M_{\odot}$ and $Z_\star \ga Z_{\odot}$), 
whereas pseudo-bulges and the central regions of disks span a wide range of values, from partial overlap 
with the former to $\sim$1 dex lower $M_\star$ and reaching down 
to subsolar metallicities of $\sim Z_{\odot}/4$.

\item Still within their central regions, bulgeless galaxies and pseudo-bulges show a clear tendency for a higher gas-phase metallicity than bulgy systems, 
a fact pointing to a substantially different chemical enrichment (and star formation) history.
This is also reflected on the stellar mass assembly histories derived in this study, indicating a 
gradual, gentler rise 
of stellar mass from bulgy towards bulgeless systems: 
at a look-back time of $\sim$6 Gyr, the percentage of the present-day stellar mass at the centers of    
bulgeless, pseudo-bulge and bulgy has been determined to be 80\%, 89\% and 93\%, respectively. 
This indicates that, whereas the stellar mass in all analyzed systems is predominantly old 
(presumably due 
to our particular selection criteria), the former two classes 
have undergone significant star-forming activity and stellar mass growth over the past few Gyrs.

\item Whereas bulge systems were found to closely obey the Faber-Jackson relation, the centers of bulgeless and 
pseudo-bulges are systematically underluminous (by up to $\sim$2.5 mag) in the r-band at a given stellar velocity dispersion 
$\sigma_*$. Given the increased importance of rotational motions in pseudo-bulges, their inferred $\sigma_*$ is rather 
an upper limit, which further underscores their kinematic departure from the Faber-Jackson relation for classical 
bulges and early-type galaxies.

\item An analysis of diagnostic emission-line ratios reveals a systematic trend along the 
bulgeless $\rightarrow$ pseudo-bulge $\rightarrow$ bulgy sequence, further highlighting the diverse 
nature of these entities: whereas BPT ratios for bulgeless galaxies are mostly compatible with those for H{\sc ii} regions,
those for pseudo-bulges suggest a broad range of gas excitation mechanisms, from star formation all the way to 
AGN. Bulgy galaxies, on the other hand, are almost exclusively found within the LINER regime of BPT diagrams.

\end{itemize}
 
Our results thus show that some trends hold for the central regions of
galaxies (from bulgeless to bulgy) concerning specific parameters
inferred from the morphological and SPS analysis.  And such relations
seem to indicate that pseudo-bulges occupy a transition region between
these two in several parameter spaces.  Namely, the fact that
pseudo-bulges seem to have residual star formation, younger stellar
populations and, consistently, a more prolonged stellar mass assembly
history than classical (bulgy) systems, lends strong support to the
notion that the central bulge-like component is forming secularly in
these galaxies.  On the other hand, the determined disk scale length
of pseudo-bulge galaxies could, in principle, be compatible with both
models (secular evolution or rapid processes such as minor mergers).
Though the majority of our results thus favor the more commonly
accepted scenario of different formation mechanisms between
pseudo-bulges and classical ones, it is far from straightforward to
exclude other possibilities, and a complex picture for the assembly of
these structures seems more reasonable as evidence builds up.

\section*{Acknowledgments}
JMG acknowledges support by Funda\c{c}\~{a}o para a Ci\^{e}ncia e a
Tecnologia (FCT) through the Fellowship SFRH/BPD/66958/2009 and
POPH/FSE (EC) by FEDER funding through the program Programa
Operacional de Factores de Competitividade (COMPETE).  PP is supported
by FCT through the Investigador FCT Contract No. IF/01220/2013 and
POPH/FSE (EC) by FEDER funding through the program COMPETE.  JMG\&PP
also acknowledge support by FCT under project
FCOMP-01-0124-FEDER-029170 (Reference FCT PTDC/FIS-AST/3214/2012),
funded by FCT-MEC (PIDDAC) and FEDER (COMPETE). This work was
supported by Funda\c c\~ao para a Ci\^encia e a Tecnologia (FCT)
through the research grant UID/FIS/04434/2013.
  
It is a pleasure to acknowledge M. Taylor for developing TOPCAT
(http://www.starlink.ac.uk/topcat/).

M. Blanton is acknowledged for support concerning the NYU-VAGC usage.

Funding for the SDSS and SDSS-II has been provided by the Alfred P. Sloan
Foundation, the Participating Institutions, the National Science Foundation,
the U.S. Department of Energy, the National Aeronautics and Space
Administration, the Japanese Monbukagakusho, the Max Planck Society, and the
Higher Education Funding Council for England. The SDSS Web Site is
http://www.sdss.org/.\\ The SDSS is managed by the Astrophysi'cal Research
Consortium for the Participating Institutions. The Participating Institutions
are the American Museum of Natural History, Astrophysical Institute Potsdam,
University of Basel, University of Cambridge, Case Western Reserve University,
University of Chicago, Drexel University, Fermilab, the Institute for Advanced
Study, the Japan Participation Group, Johns Hopkins University, the Joint
Institute for Nuclear Astrophysics, the Kavli Institute for Particle
Astrophysics and Cosmology, the Korean Scientist Group, the Chinese Academy of
Sciences (LAMOST), Los Alamos National Laboratory, the Max-Planck-Institute
for Astronomy (MPIA), the Max-Planck-Institute for Astrophysics (MPA), New
Mexico State University, Ohio State University, University of Pittsburgh,
University of Portsmouth, Princeton University, the United States Naval
Observatory, and the University of Washington.\\ This research has made use of
the NASA/IPAC Extragalactic Database (NED) which is operated by the Jet
Propulsion Laboratory, California Institute of Technology, under contract with
the National Aeronautics and Space Administration.

\bibliographystyle{mn2e}
\bibliography{apj-jour,Ribeiro15_refs}

\begin{appendix}
\section{Appendix: Tables with galaxy properties}\label{sec:appendix}

\begin{table*}
\caption{Table with general properties and structural parameters (as defined in equation \ref{multisersic}) for the pseudo-bulge galaxies studied in this paper. Columns are as follows: (1) and (2) give the J2000 coordinates of the galaxies (in degrees); (3) SDSS redshift value; (4) $g-r$ color from the NYU-VAGC catalog; (5) Total r-band magnitude of the galaxy as derived from GALFIT; (6) and (7) Effective radii, in kpc, of the structural components, disk and bulge respectively; (8) The S\'ersic index of the bulge component; (9) The morphological class assigned to the galaxy (as in Table 1).}
\begin{tabular}{c|c|c|c|c|c|c|c}
\hline
$\alpha$ (J2000) & $\delta$ (J2000) & z & $g-r$ & $M_{r,\mathrm{tot}}$ & $r_{e,d}$ & $r_{e,b}$ & $\eta_b$\\ 
\hline
212.9395 & -0.9043 & 0.0544 & 0.8190 & -20.46 &  $9.93 \pm 0.21$ & $1.79 \pm 0.02$ & $0.68 \pm 0.02$ \\
130.7137 & 52.9251 & 0.0591 & 0.8103 & -20.74 &  $6.22 \pm 0.13$ & $1.70 \pm 0.03$ & $0.62 \pm 0.04$ \\
220.8427 & 1.0981 & 0.0381 & 0.7261 & -21.03 &  $4.53 \pm 0.03$ & $1.54 \pm 0.02$ & $0.58 \pm 0.02$ \\
322.7421 & -7.0855 & 0.0278 & 0.8528 & -20.75 &  $6.23 \pm 0.05$ & $1.16 \pm 0.01$ & $0.74 \pm 0.01$ \\
30.7166 & -8.0267 & 0.0335 & 0.8086 & -20.56 &  $5.98 \pm 0.04$ & $1.11 \pm 0.01$ & $0.95 \pm 0.01$ \\
25.9110 & 13.5277 & 0.0542 & 0.8705 & -20.69 &  $14.12 \pm 0.24$ & $1.62 \pm 0.01$ & $0.46 \pm 0.02$ \\
131.1434 & 46.8706 & 0.0522 & 0.7995 & -20.09 &  $7.68 \pm 0.32$ & $2.13 \pm 0.05$ & $0.74 \pm 0.02$ \\
142.9998 & 51.3818 & 0.0333 & 0.6995 & -20.37 &  $3.91 \pm 0.07$ & $1.28 \pm 0.03$ & $1.09 \pm 0.02$ \\
194.6264 & 63.7096 & 0.0397 & 0.8460 & -19.93 &  $5.28 \pm 0.08$ & $0.97 \pm 0.01$ & $0.62 \pm 0.04$ \\
221.8257 & 58.2260 & 0.0375 & 0.7759 & -20.52 &  $3.64 \pm 0.02$ & $1.70 \pm 0.02$ & $0.42 \pm 0.02$ \\
254.4245 & 33.9194 & 0.0597 & 0.8135 & -19.88 &  $9.36 \pm 0.20$ & $1.57 \pm 0.02$ & $1.01 \pm 0.03$ \\
117.8212 & 32.7403 & 0.0557 & 0.7645 & -20.72 &  $5.16 \pm 0.06$ & $1.39 \pm 0.02$ & $0.34 \pm 0.04$ \\
155.1420 & 7.8518 & 0.0441 & 0.8683 & -20.31 &  $5.47 \pm 0.08$ & $1.04 \pm 0.01$ & $0.81 \pm 0.02$ \\
162.9716 & 8.8632 & 0.0523 & 0.7222 & -20.75 &  $7.20 \pm 0.13$ & $1.86 \pm 0.03$ & $0.80 \pm 0.03$ \\
231.4910 & 48.2958 & 0.0361 & 0.7062 & -20.35 &  $4.42 \pm 0.03$ & $2.22 \pm 0.03$ & $1.46 \pm 0.01$ \\
145.2456 & 40.0365 & 0.0412 & 0.8357 & -20.02 &  $5.40 \pm 0.07$ & $0.81 \pm 0.01$ & $0.33 \pm 0.02$ \\
158.9306 & 12.2550 & 0.0495 & 0.8860 & -20.78 &  $8.84 \pm 0.16$ & $1.96 \pm 0.02$ & $0.73 \pm 0.02$ \\
188.8515 & 47.6891 & 0.0454 & 0.9074 & -20.15 &  $4.58 \pm 0.08$ & $0.92 \pm 0.01$ & $0.86 \pm 0.02$ \\
253.8737 & 23.3854 & 0.0553 & 0.9055 & -21.29 &  $10.46 \pm 0.07$ & $1.12 \pm 0.02$ & $0.70 \pm 0.06$ \\
237.9711 & 27.2427 & 0.0589 & 0.7937 & -20.72 &  $6.11 \pm 0.09$ & $1.43 \pm 0.01$ & $0.42 \pm 0.02$ \\
19.6423 & -0.2283 & 0.0472 & 0.7822 & -20.85 &  $5.90 \pm 0.05$ & $1.38 \pm 0.01$ & $0.35 \pm 0.01$ \\
18.7512 & 0.0309 & 0.0497 & 0.7580 & -20.54 &  $5.22 \pm 0.09$ & $0.96 \pm 0.01$ & $0.86 \pm 0.02$ \\
197.9124 & 34.6365 & 0.0374 & 0.8359 & -20.88 &  $5.35 \pm 0.04$ & $5.49 \pm 0.06$ & $1.21 \pm 0.01$ \\
151.1738 & 28.3567 & 0.0516 & 0.9225 & -20.83 &  $7.58 \pm 0.06$ & $0.88 \pm 0.01$ & $0.54 \pm 0.03$ \\
135.5952 & 14.5252 & 0.0301 & 0.8168 & -21.27 &  $14.23 \pm 0.16$ & $2.66 \pm 0.01$ & $0.82 \pm 0.01$ \\
168.9758 & 20.7440 & 0.0599 & 0.7755 & -20.80 &  $8.71 \pm 0.11$ & $1.97 \pm 0.01$ & $1.00 \pm 0.01$ \\
174.4652 & 21.9742 & 0.0303 & 0.7217 & -21.96 &  $12.86 \pm 0.13$ & $1.71 \pm 0.00$ & $1.31 \pm 0.01$ \\
223.5155 & 18.4004 & 0.0571 & 0.8136 & -20.61 &  $5.12 \pm 0.06$ & $1.65 \pm 0.02$ & $0.73 \pm 0.03$ \\
134.9453 & -0.0056 & 0.0527 & 0.7735 & -21.70 &  $8.45 \pm 0.07$ & $0.88 \pm 0.01$ & $1.45 \pm 0.02$ \\
\hline
\end{tabular}
\label{tab_allgals}
\end{table*}

\begin{table*}
\caption{Same as table \ref{tab_allgals} but for the Bulgeless class galaxies.}
\label{tab_bugeless}
\begin{tabular}{c|c|c|c|c|c|c|c}
\hline
$\alpha$ (J2000) & $\delta$ (J2000) & z & $g-r$ & $M_{r,\mathrm{tot}}$ & $r_{e,d}$ & $r_{e,b}$ & $\eta_b$\\ 
\hline
212.3728 & 0.1437 & 0.0541 & 0.7417 & -20.02 &  $3.59 \pm 0.03$ & - & -  \\
15.7648 & 13.4973 & 0.0578 & 0.7104 & -20.12 &  $3.35 \pm 0.02$ & - & -  \\
256.8101 & 65.3668 & 0.0537 & 0.7276 & -20.58 &  $3.20 \pm 0.02$ & - & -  \\
121.1723 & 45.7877 & 0.0504 & 0.7333 & -20.82 &  $3.66 \pm 0.02$ & - & -  \\
129.1653 & 47.2543 & 0.0528 & 0.8464 & -19.72 &  $3.27 \pm 0.03$ & - & -  \\
334.8222 & -1.1872 & 0.0571 & 0.7529 & -19.88 &  $2.58 \pm 0.03$ & - & -  \\
52.5906 & 0.2631 & 0.0373 & 1.1631 & -20.34 &  $4.14 \pm 0.02$ & - & -  \\
7.5758 & -0.5022 & 0.0585 & 0.7055 & -20.64 &  $6.32 \pm 0.07$ & - & -  \\
233.1491 & 49.3842 & 0.0521 & 0.7849 & -19.89 &  $2.89 \pm 0.02$ & - & -  \\
257.1751 & 28.5217 & 0.0451 & 0.7109 & -20.92 &  $4.61 \pm 0.01$ & - & -  \\
209.7411 & 58.2354 & 0.0593 & 0.7557 & -20.96 &  $2.33 \pm 0.01$ & - & -  \\
242.0544 & 44.1528 & 0.0492 & 0.7436 & -20.15 &  $4.79 \pm 0.04$ & - & -  \\
134.1697 & 5.8765 & 0.0592 & 0.7580 & -17.81 &  $3.93 \pm 0.05$ & - & -  \\
145.5350 & 9.7321 & 0.0589 & 0.7888 & -19.88 &  $3.26 \pm 0.02$ & - & -  \\
127.3326 & 6.2958 & 0.0484 & 0.7508 & -20.92 &  $2.97 \pm 0.01$ & - & -  \\
158.5929 & 44.3970 & 0.0522 & 0.7227 & -20.36 &  $2.09 \pm 0.01$ & - & -  \\
204.2524 & 43.4256 & 0.0436 & 0.7958 & -20.42 &  $7.24 \pm 0.05$ & - & -  \\
166.6466 & 44.0469 & 0.0366 & 0.7460 & -20.04 &  $2.25 \pm 0.01$ & - & -  \\
197.1282 & 50.6423 & 0.0293 & 0.7530 & -20.46 &  $2.13 \pm 0.00$ & - & -  \\
199.2471 & 7.7240 & 0.0487 & 0.7647 & -20.26 &  $4.07 \pm 0.03$ & - & -  \\
196.6814 & 9.6532 & 0.0565 & 0.7108 & -20.42 &  $2.29 \pm 0.01$ & - & -  \\
204.0016 & 6.5261 & 0.0231 & 0.7221 & -20.01 &  $2.80 \pm 0.01$ & - & -  \\
243.0000 & 30.0477 & 0.0482 & 0.7935 & -20.50 &  $5.30 \pm 0.04$ & - & -  \\
111.0129 & 40.8093 & 0.0497 & 0.9650 & -20.06 &  $4.39 \pm 0.06$ & - & -  \\
181.4486 & 33.8394 & 0.0539 & 0.7074 & -20.35 &  $3.74 \pm 0.05$ & - & -  \\
249.7326 & 13.3908 & 0.0507 & 0.7658 & -20.41 &  $3.83 \pm 0.03$ & - & -  \\
147.9439 & 27.5461 & 0.0330 & 0.7502 & -20.52 &  $2.24 \pm 0.01$ & - & -  \\
193.9952 & 30.3637 & 0.0512 & 0.7330 & -20.77 &  $4.28 \pm 0.02$ & - & -  \\
159.7373 & 25.7561 & 0.0510 & 1.1405 & -20.12 &  $4.81 \pm 0.06$ & - & -  \\
179.4989 & 25.1587 & 0.0580 & 0.7424 & -19.97 &  $2.00 \pm 0.01$ & - & -  \\
120.5747 & 11.4264 & 0.0600 & 0.8377 & -19.32 &  $2.00 \pm 0.01$ & - & -  \\
153.5945 & 18.4474 & 0.0441 & 0.8836 & -19.58 &  $2.17 \pm 0.01$ & - & -  \\
148.8226 & 17.6874 & 0.0447 & 0.8021 & -20.43 &  $2.47 \pm 0.01$ & - & -  \\
159.9893 & 17.6721 & 0.0571 & 0.6974 & -20.30 &  $4.05 \pm 0.04$ & - & -  \\
171.8496 & 19.6471 & 0.0520 & 0.7675 & -19.93 &  $2.22 \pm 0.01$ & - & -  \\
211.4471 & 15.1940 & 0.0595 & 0.7129 & -20.72 &  $3.17 \pm 0.03$ & - & -  \\
244.2731 & 11.4185 & 0.0402 & 0.7123 & -20.63 &  $3.46 \pm 0.01$ & - & -  \\
165.8052 & 7.7149 & 0.0554 & 0.7294 & -20.93 &  $3.35 \pm 0.02$ & - & -  \\
\hline
\end{tabular}
\end{table*}

\begin{table*}
\caption{Same as table\ref{tab_allgals} but for the Bulgy class galaxies.}
\label{tab_bulgy}
\begin{tabular}{c|c|c|c|c|c|c|c}
\hline
$\alpha$ (J2000) & $\delta$ (J2000) & z & $g-r$ & $M_{r,\mathrm{tot}}$ & $r_{e,d}$ & $r_{e,b}$ & $\eta_b$\\ 
\hline
55.9895 & 0.4378 & 0.0400 & 0.7490 & -21.17 &  - & $4.16 \pm 0.07$ & $5.22 \pm 0.06$ \\
57.6747 & 1.0408 & 0.0371 & 0.7447 & -21.13 &  - & $3.62 \pm 0.07$ & $5.56 \pm 0.06$ \\
241.4743 & -0.5504 & 0.0560 & 0.7526 & -21.09 &  - & $3.35 \pm 0.08$ & $5.05 \pm 0.08$ \\
241.2225 & -0.0475 & 0.0520 & 0.7514 & -20.86 &  - & $2.69 \pm 0.06$ & $4.80 \pm 0.09$ \\
243.0857 & 0.8049 & 0.0578 & 0.7354 & -21.49 &  - & $6.99 \pm 0.35$ & $6.16 \pm 0.15$ \\
191.5058 & -1.0732 & 0.0474 & 0.7499 & -20.89 &  - & $3.64 \pm 0.11$ & $5.56 \pm 0.11$ \\
239.8879 & -1.0989 & 0.0553 & 0.7294 & -20.92 &  - & $2.72 \pm 0.05$ & $3.68 \pm 0.06$ \\
211.4747 & -0.7454 & 0.0593 & 0.7498 & -21.78 &  - & $4.44 \pm 0.08$ & $5.07 \pm 0.07$ \\
226.1841 & -0.3520 & 0.0548 & 0.7363 & -21.98 &  - & $7.30 \pm 0.14$ & $4.83 \pm 0.05$ \\
244.4877 & -0.3815 & 0.0572 & 0.7142 & -21.14 &  - & $4.05 \pm 0.16$ & $6.41 \pm 0.14$ \\
172.0247 & 0.1322 & 0.0495 & 0.7483 & -20.25 &  - & $1.75 \pm 0.02$ & $3.34 \pm 0.07$ \\
217.4984 & 0.2003 & 0.0554 & 0.7196 & -20.82 &  - & $2.20 \pm 0.05$ & $4.81 \pm 0.10$ \\
223.4544 & 0.0897 & 0.0439 & 0.7667 & -21.20 &  - & $3.60 \pm 0.09$ & $6.52 \pm 0.09$ \\
239.1147 & 0.0593 & 0.0397 & 0.7437 & -20.39 &  - & $4.58 \pm 0.18$ & $5.92 \pm 0.12$ \\
247.5797 & 0.1850 & 0.0582 & 0.7267 & -20.79 &  - & $3.29 \pm 0.11$ & $5.11 \pm 0.13$ \\
172.6222 & 0.4943 & 0.0290 & 0.7576 & -21.10 &  - & $6.26 \pm 0.16$ & $5.62 \pm 0.07$ \\
204.2834 & 0.4521 & 0.0479 & 0.7558 & -21.19 &  - & $3.83 \pm 0.07$ & $5.22 \pm 0.06$ \\
222.5090 & 0.5788 & 0.0404 & 0.7378 & -21.59 &  - & $4.99 \pm 0.14$ & $7.84 \pm 0.10$ \\
182.7670 & 0.9723 & 0.0204 & 0.7531 & -22.33 &  - & $22.11 \pm 0.47$ & $6.10 \pm 0.04$ \\
153.4382 & -0.8772 & 0.0418 & 0.7437 & -20.12 &  - & $4.67 \pm 0.23$ & $5.36 \pm 0.14$ \\
242.5503 & 0.7837 & 0.0433 & 0.7659 & -22.17 &  - & $7.77 \pm 0.17$ & $5.48 \pm 0.09$ \\
210.9270 & -1.1373 & 0.0269 & 0.7541 & -20.34 &  - & $2.75 \pm 0.03$ & $4.50 \pm 0.11$ \\
213.7526 & -0.7970 & 0.0385 & 0.7543 & -22.15 &  - & $5.01 \pm 0.03$ & $3.98 \pm 0.02$ \\
214.3278 & 0.0946 & 0.0526 & 0.7821 & -21.80 &  - & $8.51 \pm 0.22$ & $5.32 \pm 0.29$ \\
222.7663 & -0.4622 & 0.0432 & 0.7669 & -21.28 &  - & $5.32 \pm 0.08$ & $4.11 \pm 0.31$ \\
193.8191 & 0.2469 & 0.0476 & 0.7537 & -22.63 &  - & $9.18 \pm 0.13$ & $5.83 \pm 0.09$ \\
\hline
\end{tabular}
\end{table*}

\begin{table*}
\caption{Same as table \ref{tab_allgals} but for the Intermediate-$\eta$ bulge class galaxies.}
\label{tab_intermediate}
\begin{tabular}{c|c|c|c|c|c|c|c}
\hline
$\alpha$ (J2000) & $\delta$ (J2000) & z & $g-r$ & $M_{r,\mathrm{tot}}$ & $r_{e,d}$ & $r_{e,b}$ & $\eta_b$\\ 
\hline
211.8419 & -1.0964 & 0.0551 & 0.7439 & -21.43 &  $7.82 \pm 0.06$ & $0.76 \pm 0.02$ & $1.52 \pm 0.12$  \\
227.7024 & -1.0625 & 0.0542 & 0.7742 & -21.64 &  $8.23 \pm 0.10$ & $1.17 \pm 0.01$ & $1.51 \pm 0.02$  \\
223.0669 & -0.2936 & 0.0434 & 0.7345 & -21.76 &  $7.09 \pm 0.10$ & $2.61 \pm 0.12$ & $3.26 \pm 0.07$  \\
173.8470 & 0.0906 & 0.0292 & 0.6973 & -20.31 &  $6.47 \pm 0.22$ & $1.31 \pm 0.04$ & $3.16 \pm 0.06$  \\
200.4663 & 0.1394 & 0.0347 & 0.7524 & -21.21 &  $5.78 \pm 0.03$ & $0.89 \pm 0.02$ & $2.89 \pm 0.04$  \\
211.0308 & 0.1440 & 0.0477 & 0.7341 & -21.30 &  $5.50 \pm 0.03$ & $0.54 \pm 0.02$ & $3.48 \pm 0.50$  \\
226.1676 & 0.6045 & 0.0402 & 0.7644 & -21.56 &  $7.67 \pm 0.05$ & $0.93 \pm 0.01$ & $1.58 \pm 0.02$  \\
169.8617 & 0.9681 & 0.0399 & 0.7654 & -21.13 &  $4.58 \pm 0.04$ & $0.58 \pm 0.01$ & $1.66 \pm 0.04$  \\
190.7771 & -0.4379 & 0.0473 & 0.8125 & -22.33 &  $8.94 \pm 0.07$ & $1.62 \pm 0.05$ & $2.27 \pm 0.06$  \\
167.1651 & 0.2843 & 0.0249 & 0.7578 & -20.37 &  $3.66 \pm 0.02$ & $0.72 \pm 0.01$ & $2.56 \pm 0.03$  \\
189.7547 & 0.3656 & 0.0230 & 0.7874 & -21.95 &  $9.02 \pm 0.06$ & $1.50 \pm 0.01$ & $2.49 \pm 0.01$  \\
163.6087 & 0.6637 & 0.0374 & 0.7811 & -21.24 &  $6.78 \pm 0.07$ & $1.12 \pm 0.02$ & $2.75 \pm 0.05$  \\
21.8550 & 14.0545 & 0.0236 & 0.7966 & -20.60 &  $5.18 \pm 0.10$ & $1.22 \pm 0.03$ & $2.40 \pm 0.04$  \\
\hline
\end{tabular}
\end{table*}

\end{appendix}

\label{lastpage}

\end{document}